\documentclass[pra,twocolumn,superscriptaddress,amsmath,amssymb,aps,floatfix]{revtex4-2}
\usepackage{dcolumn}% Align table columns on decimal point
\usepackage{bm}% bold math
\usepackage{amsmath,amssymb}
\usepackage{algorithm}
\usepackage{algpseudocode}
\usepackage{graphics}
\usepackage{amsfonts}
\usepackage{epsfig}
\usepackage{float}
\usepackage{hyperref}
\usepackage{adjustbox}
\usepackage{rotating}
\usepackage{float}
\usepackage{dblfloatfix}
\usepackage{blindtext}
\usepackage{multirow}
\usepackage{graphicx}
\usepackage{dcolumn}
\usepackage{makecell}
\usepackage{bbold}
\usepackage{bm}
\usepackage{amsmath}
\usepackage{import}
\usepackage{physics}
\usepackage{verbatim}
\usepackage{listings}
\usepackage{xcolor}
\usepackage[table]{xcolor}
\usepackage{xr-hyper}
\usepackage{hyperref}
\usepackage{cleveref}
\usepackage{newfloat}
\DeclareFloatingEnvironment[name={Fig S},fileext=lof]{suppfigure}
\begin{document}
\title{Quantum walks reveal topological flat bands, robust edge states and topological phase transitions in cyclic graphs}
\author{Dinesh Kumar Panda}
\email{dineshkumar.quantum@gmail.com}
\author{Colin Benjamin}
\email{colin.nano@gmail.com}
\affiliation{School of Physical Sciences, National Institute of Science Education and Research, Bhubaneswar, Jatni 752050, India}
\affiliation{Homi Bhabha National Institute, Training School Complex, Anushaktinagar, Mumbai
400094, India}

\begin{abstract}
Topological phases, edge states, and flat bands in synthetic quantum systems are a key resource for topological quantum computing and noise-resilient information processing. We introduce a scheme based on step-dependent quantum walks on cyclic graphs, termed cyclic quantum walks (CQWs), to simulate exotic topological phenomena using discrete Fourier transforms and an effective Hamiltonian. Our approach enables the generation of both gapped and gapless topological phases, including Dirac cone-like energy dispersions, topologically nontrivial flat bands, and protected edge states, all without resorting to resource-intensive split-step or split-coin quantum walk protocols. Odd and even-site cyclic graphs exhibit markedly different spectral characteristics, with rotationally symmetric flat bands emerging exclusively in $4n$-site graphs ($n\in \mathbf{N}$). We analytically establish the conditions for the emergence of topological, gapped flat bands and show that gap closings in rotation space imply the formation of Dirac cones in momentum space. Further, we engineer protected edge states at the interface between distinct topological phases in both odd and even cycle graphs. We numerically demonstrate that the edge states are robust against moderate static and dynamic gate disorder as well as remain stable against phase-preserving perturbations and are independent of initial states. This scheme serves as a resource-efficient and versatile platform to engineer topological phases, transitions, edge states, and flat bands in small-scale quantum systems, opening new avenues for robust quantum memory, protected state transfer, and compact implementations of fault-tolerant quantum technologies.
\end{abstract}

\maketitle
\newpage
\twocolumngrid

\textcolor{brown}{Introduction.--} Topological phases, edge states, and flat bands lie at the heart of contemporary research in condensed matter physics and topological quantum computing (TQC)~\cite{pan16,pan23,pxue}. This originated with the discovery of integer quantum Hall effect~\cite{hall1,hall2} and has accelerated significantly through theoretical prediction~\cite{kita21,kita22,kita23} and experimental realization~\cite{kita24,kita25} of topological insulators and fractional charges~\cite{kita25,kita26}. 
Topological phases typically emerge in systems with gapped energy bands, while edge states arise at
the interface of these phases~\cite{pan-1,pan-2,pxue}. Energy bands can close their gap in distinct ways: Dirac cone (linear closing in momentum), Fermi arc (nonlinear closing in momentum), flat bands (energy constant in momentum). Gapped flat bands are easier to isolate and find applications in strong correlations~\cite{correl}, Mott phases~\cite{mott}, fractional quantum Hall states~\cite{fqh}, while gapless flat bands can host critical states or semimetallic behaviours of matter and have applications in quantum critical systems, semimetals, and exotic transport~\cite{fqh,correl,mott, semimetal,criticalsyst}. Topological phases have been realized experimentally in physical systems, e.g., with photons~\cite{kita28,pxue}, ultracold atoms or molecules~\cite{kita42,kita40}. offer a promising platform for topological qubits, fault-tolerant TQC and topological quantum information (QI) processing, where fault tolerance arises from nonlocal encoding of quasiparticle states, rendering them intrinsically resilient to errors from local perturbations~\cite{kitaev2003, nayak2008, stern2010}. Topological edge states can be used for robust quantum information storage in quantum memory devices, and for noise-resilient state transfer in quantum communication networks~\cite{qmemory, qmemory2, edge1, edge2, edge4}.

Unfortunately, the number of material-based topological insulators is small, and topological properties, e.g., quantized-edge conductance, symmetry-protected modes, and bulk-boundary correspondence, are constrained to specific material classes and symmetry conditions~\cite{Schnyder2008,Zhang2019,Hasan2010,Halperin1982, chandra-topo}. This drives researchers to find ways to create synthetic quantum systems capable of hosting nontrivial topological phases. Among various approaches, discrete-time quantum walks (QWs) have emerged as a powerful framework for generating such phases. QWs describe time-evolution of quantum particles having internal states on discrete lattices, where interference, coherence and entanglement govern the dynamics~\cite{Aharonov,pxue,ina2023,joshua2025,Cinthia,portugal,p4}. Recent works report that QW on 1D/2D/3D lattice and with photons can simulate a range of topological phenomena~\cite{pan15,pan16, pan17, pan21,pan22,pan23,pxue,karimi19,pan-1,pan-2, kita-pra,asboth,barry-pra}. Owing to their tunability and compatibility with various physical architectures, QWs can offer an attractive route to realize and explore topological phases, especially in regimes difficult or impossible to access in condensed matter systems.

However, an attempt to simulate topological phenomena, flat bands and edge states using QW on cyclic graphs has been missing, although Ref.~\cite{karimi19} presents a report restricted to calculating the Zak phase of a QW with Hadamard gate (coin) on a 6-site cyclic graph. Notably, QW dynamics on cyclic graphs (i.e., cyclic quantum walk or CQW) describes the wave-packet dynamics of single particles and can effectively simulate complex quantum phenomena, including coherent energy transport and quantum interference effects in ring-structured systems~\cite{karimi19,bian}. Further, CQWs are less resource-consuming to implement experimentally due to the finite size working Hilbert space~\cite{bian,karimi19}. This allows a more resource-efficient implementation of simulation of topological effects via cyclic graphs in real physical setups, in contrast to 1D/2D/3D lattice. As proposed by us in \cite{pral}, a CQW is more resource-saving than 1D/2D/3D lattice QWs, when used in quantum cryptography for message encryption-decryption and quantum direct communication. In this work, we harness the potential of CQW on finite odd-even cyclic graphs, to simulate exotic topological effects. We demonstrate CQWs serve as highly flexible and most resource-saving platforms to generate diverse energy dispersion (band structures), Dirac cones (band closing), topological phases (nonzero winding numbers), topological gapped flat bands and topologically protected edge states (we also show their robustness against moderate static-dynamic disorder, change in initial states, and perturbations) in real quantum systems. These edge states can enable superior resource-efficiency in quantum information storage and noise-resilient state transfer, using small-scale quantum architectures and experimental implementation with minimal
resource overhead. Our framework offers fine-grained control over these topological features through step-dependency parameter, site number $N$, periodic evolution (unique to cyclic graphs) and coin-rotation angles.

 Below, we introduce the theory of CQW dynamics and the physics of evaluating energy band structures, group velocity and effective mass via stroboscopic evolution. The procedure to derive topological invariants in cyclic graphs and results on topological phases, transitions and flat bands are established. The design of topologically protected edge states in odd and even cyclic graphs is demonstrated, along with their robustness against static-dynamic disorder and perturbations and a photon-based experimental implementation. Additional derivations and results are provided in Supplementary Material (SM)~\cite{sm} and our Python code to design state-independent robust edge states is on GitHub~\cite{git}.

\textcolor{brown}{Model.--}
CQW describes the propagation of the spatial distribution of a single quantum particle (e.g., electron or photon)
on an $N$-cycle graph (e.g., atomic sites, orbital angular momentum). CQW evolution has spatial symmetry and can be diagonalized via Fourier transform (FT) methods, i.e, $
        |k'\rangle = \frac{1}{\sqrt{N}} \sum_{x=0}^{N-1} e^{i\frac{2\pi}{N}k'x} |x\rangle_p,$
where $k',x \in [0, N-1]$. The walker's motion at any time step $t$ can be clockwise or anticlockwise, which is governed by a translation/shift operator $\hat{S} = \sum_{q=0}^{1}\sum_{x=0}^{N-1}\ket{(x+(-1)^{q}) \text{ mod } N}_p\bra{x}_p\otimes\ket{q}_c\bra{q}_c$ contingent upon the action of a single-qubit gate (coin) $\hat{C}_{2}(\theta, T) \equiv e^{-i \frac{T\theta}{2} \sigma_y}$.  The time-evolution of the quantum walker at a particular time-step is given as, \begin{equation}
U_{N}= \hat{S}.[I_N\otimes \hat{C}_{2}(\theta, T)]\;,
\label{eq2}
\end{equation}
\noindent
and the evolved quantum state of the walker is, $\ket{\psi(t)} = (U_{N})^t \ket{\psi(0)}$.
%and the evolved quantum state $\ket{\psi(t)} = U_{k}(t)U_{k}(t-1)...U_{k}(1) \ket{\psi(0)}$
\noindent
On diagonalizing $U_N$ in quasi-momentum $k'$-basis and transforming it to a stroboscopic evolution via an effective Hamiltonian ($\hat{H}$ in units of $\hbar=1$),
\begin{equation}
U_N = e^{-i H},\;\hat{H} = E(k) \hat{n}(k) \cdot \vec{\sigma},\; k=\frac{2\pi k'}{N},
\label{eq5}
\end{equation}
 where $\hat{n}$ is the winding vector in Bloch sphere, we evaluate (see SM Sec.~A) the energy dispersion relation for an arbitrary $N$-cycle graph,
\begin{equation}
\resizebox{0.89\linewidth}{!}{$
E(k) = \pm \cos^{-1}(\cos k \cos \frac{T\theta}{2})= \pm \cos^{-1}(\cos \frac{2\pi k'}{N} \cos \frac{T\theta}{2}).
    \label{Ekcqw}
    $}
\end{equation}
\begin{figure}[h]
\includegraphics[width = 8cm,height=2.2cm]{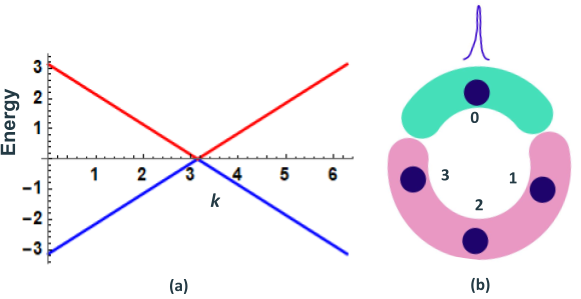}
\caption{Schematics of (a) a Dirac cone for CQW, where energy gap closing is linear; (b) two distinct topological phase regimes are shown in green and red on a 4-cycle and an edge state (wave peak in blue) is expected to form at the boundary between the phases.}
% and a nonlinear gap closing refers to Fermi arc
\label{f0}
\end{figure}
%dramatic amplification of 
% Topological phase transitions occur at and across the boundary states.
We evaluate the group velocity and effective mass of the particle arising due to the curvature of energy bands, as in SM Sec.~A, and both can be controlled via $\theta,N,T$; such control has importance in determining flat band formation (SM Sec.~B), carrier mobility, diffusion rates, and wave packet spreading in solid-state quantum systems~\cite{effmass1,effmass2,effmass3}.
%Such control has potential use in advancing modern electronics, optoelectronics, and quantum technologies as in solid-state systems, the effective mass plays a key role in determining carrier mobility, diffusion rates, and wave packet spreading~\cite{effmass1,effmass2,effmass3} and a low effective mass corresponds to faster spreading, lower inertia, and higher mobility. 
We derive the topological invariant~\cite{karimi19,filippo17,panahiyan20}, winding number $\omega_{\theta,T,N}$, to characterize the topological phases of CQW on $N$-cycle graphs, using an arbitrary coin $\hat{C}_{2}(\theta, T)$, see SM Sec.~A, i.e.,
\begin{equation}
\resizebox{0.82\linewidth}{!}{$
  \omega_{\theta,T,N}=
\sum_{k'=0}^{N-1}\frac{ \sin[\frac{T\theta}{2}]}{N(1-\cos\left[\frac{2\pi k'}{N}\right]^2 \cos[\frac{T\theta}{2}]^2)}.
\label{wind1t}
$}
\end{equation}

For example, with Hadamard coin ($\theta=\frac{\pi}{2}, T=1$), for $N=7$ (7-cycle) we get $\omega_{\frac{\pi}{2},1,7} \approx 1.00001$ and  for $N=8$ (8-cycle), $\omega_{\frac{\pi}{2},1,8} \approx 1.00173$, and when $N=1000$ (finite large $N$ limit), $\omega_{\frac{\pi}{2},1,1000}=1$. A nonzero (zero) winding number indicates a topological (trivial) phase of the quantum systems evolving via CQW dynamics, and for small cyclic graphs, it is reasonable to study topological effects as these are less resource-intensive and more feasible to generate experimentally than a 1D infinite line or 2D/3D lattices, see more details in SM Secs.~A and F.  Thus, exploiting CQW on finite lattices will help simulate topological phases, band closing and edge states in physical systems in a resource-saving manner in experiments, say with photonic or ion trap circuits~\cite{bian, karimi19}. Below, we show that step-dependent ($T\ge 2$) and step-independent ($T=1$) CQW dynamics offer excellent control over topological features such as edge states, flat bands, Dirac cones and topological phase transitions, via rotation angles, site number and step-dependence ($T$) on finite-size cyclic graphs.

%Thus, exploiting CQW on finite lattices will help simulate topological phases, band closing and edge states in physical systems in a resource-saving manner in experiments, say with photonic or ion trap circuits~\cite{bian, karimi19}. Below, we show that step-dependent ($T\ge 2$) and step-independent ($T=1$) CQW dynamics offer excellent control over topological features such as edge states, flat bands, Dirac cones and topological phase transitions, via rotation angles, site number and step-dependence ($T$) on finite-size cyclic graphs.}

\begin{widetext}  
 
\begin{figure}[H]
\includegraphics[width = 18cm,height=6.2cm]{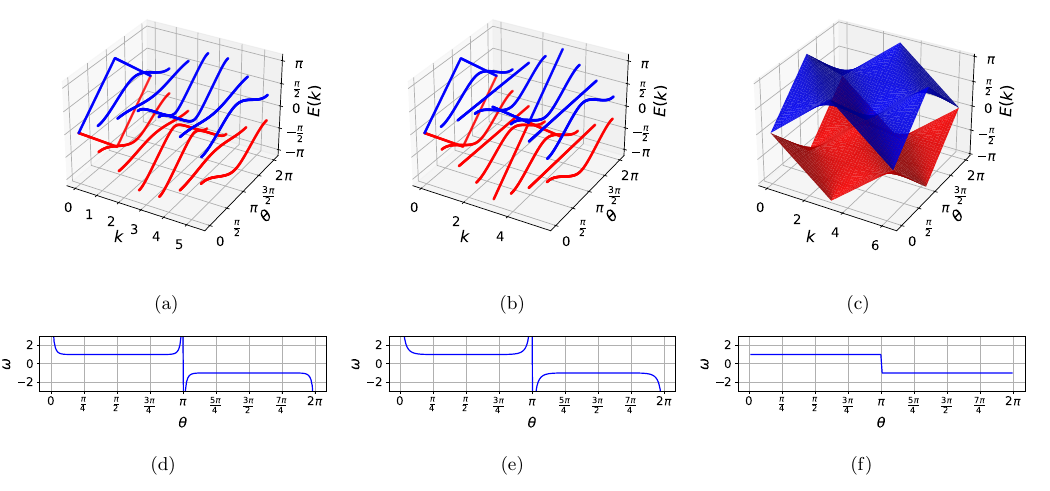}
\caption{Energy dispersion vs quasi-momenta $k$ and rotation angle $\theta$ for (a) $N=7$, (b) $N=8$-cycles and (c) $N=1000$ (with Dirac cones). The blue (red) surface refers to the upper (lower) energy band. Winding number $\omega$ vs $\theta$ for (d) $N=7$, (e) $N=8$ and (f) $N=1000$ (continuum limit), for step-dependent ($T=2$) CQW. }
\label{f78t2}
\end{figure}

\end{widetext}

\textcolor{brown}{Energy dispersion and topological phases.--} The energy dispersion and winding number ($\omega$), see Eqs.~(\ref{Ekcqw})-(\ref{wind1t}), are plotted in Fig.~\ref{f78t2} for step-dependent coin ($T=2$) and in Fig.~3 of SM Sec.~A for step-independent coin ($T=1$) for $N=7,8$-cycles, see SM Sec.~A for 3 and 4-cycles and also 7,8-cycles with higher step-dependency, e.g., $T\ge3$. Notably, the trend in $\omega$ vs. $\theta$ for odd and even cycles is identical, and the odd-even distinction vanishes as $N\rightarrow$ large. However, in finite cycles (e.g., $N=3,4,7,8$), odd-even distinction is relevant for energy dispersion, band-closing and flat bands.
Further, we observe that with increasing $T$, the number of locations of energy gap closing (Dirac cones) increases, and so does the variety of winding numbers, see Fig.~\ref{f78t2} in comparison to Fig.~\ref{f78t1}. Thus, step-dependent coins ($T>1$) show a larger number of distinct topological phases (with topological phase transitions) than the step-independent case (always $\omega=1$ with no phase transition).

One distinct feature in even 8-cycle (or, 4-cycle) as compared to odd 7-cycle (or, 3-cycle) is that the number of band-closing locations is larger in 8-cycle (or, 4-cycle). Besides, we see band closing beyond trivial $k=0$, e.g., at $k=\pi$ only in 8-cycle (4-cycle) for particular coins ($\theta$), and it holds for both step-independent and step-dependent CQW regardless of $T$. 
%Further, we observe that with increasing $T$, the number of locations where the energy band-gap closes increases, and so does the number of edge states for small and very large ($N=1000$)-cycles. The increase in the number of edge states is evident from the increased variation of nonzero winding numbers, with increasing $T$, see Fig.~\ref{f78t2} in comparison to Fig.~\ref{f78t1}. Herein, a nonzero (zero) winding number indicates a topological (trivial) phase of the quantum systems evolving via CQW dynamics, and for step-dependent coins ($T>1$) shows a larger number of distinct topological phases (more varieties of winding numbers) than the step-independent coins ($T=1$ case).

\begin{widetext}

\begin{figure}[H]
\includegraphics[width = 18cm,height=3cm]{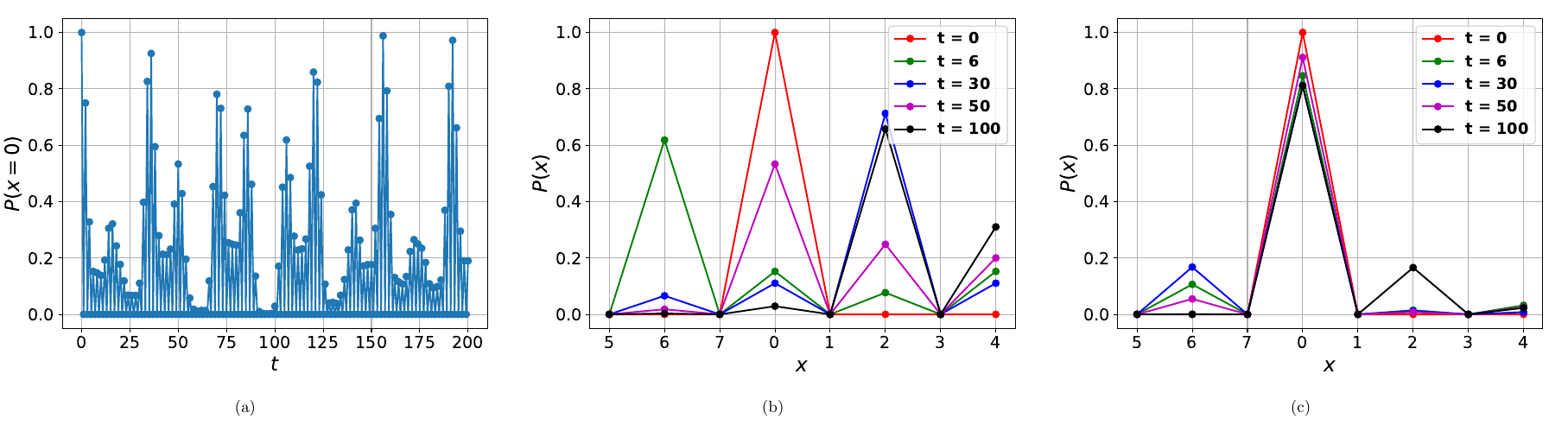}
\caption{(a) Probability of the particle at position $x=0$ vs time-step $t$ showing chaotic evolution; (b) Absence of edge state due to identical topological phase ($\omega=1$) throughout position space i.e., no boundary; (c) Generation of edge state (persistent over $t$) at the interface (site 0) between two distinct phases (i.e., with $\omega=-1$ and $\omega=+1$), via step-dependent CQW ($T=2$), for 8-cycle.}
\label{edge-8}
\end{figure} 

\end{widetext}

We also analytically show (in SM Sec.~B) that the gap closing (Dirac cones) at $E(k)=0$ happens under the condition: $\theta \in \{0,\frac{4 j\pi}{T} :\; k=0, 2\pi \;(\text{or, } k' = 0, N) \}$ and $\{\frac{(4j+2)\pi}{T}:\;  k=\pi\;(\text{or, }k' = \frac{N}{2})\}, j\in \mathbf{Z_{+}}$, which allows one to control gap closing and Dirac cone locations with the CQW parameters: $\theta, T, N$. 
 This analysis is supported by Figs.~\ref{f78t2}-\ref{f78t1} for gap closing (Dirac cones) for the derived $\theta$ values, see also 2D plots in SM Sec.~B. The gap closing in rotation angle $\theta$ also implies Dirac cones (linear gap-closing in momentum space $k$), see SM Sec.~B and Fig.~\ref{f78t2}. Thus, one can control the conducting phases of the CQW system using only the coin operators. For instance, in Fig.~\ref{f78t2}(a-b) with $k=0, \theta=0$ for 7-cycle and with $k=0, \theta=0$ or $k=\pi, \theta=2\pi$ for 8-cycle, we see gap closing in $\theta$ and these $\theta$ values show linear energy-gap closing in $k$ (continuum limit) too, see Fig.~\ref{f78t2}(c).

Further, we have analytically derived the condition for flat bands, i.e., zero group velocity and undefined effective mass: $\theta=(2n+1)\frac{\pi}{T}, \;n\in \mathbb{Z_+}\cup \{0\}$, see SM Sec.~B. For instance, $\theta=\frac{\pi}{2},\frac{3\pi}{2}$ with $T=2$ or $\theta=\pi$ with $T=1$ lead to the appearance of gapped flat bands (which are topological) at $E(k)=\pm \frac{\pi}{2}$, see Figs.~\ref{f78t2}(c), ~\ref{f78t1}(c). Notably, gapless flat bands are not possible in CQW. Moreover, we prove that rotational flat bands (dispersion is independent of rotation angle $\theta$ or coin, i.e., flat band with rotation angle) manifest only for even cycles with $N$ being a multiple of 4, i.e., only for $N=4,8,...$-cycles and are absent for all odd-cycles and all their even multiples, i.e., $N=3,5,7,6,...$, see Fig.~\ref{f78t2}. This holds for all $T$, and further details and examples are mentioned in SM Secs.~A, B.

%\textcolor{red}{We can analytically show where band closes with $E=0,\pm \pi$ in 3-cycle and 4-cycle as a fucntion of $\theta, N$--ONGOING}. 
%\end{widetext}

\textcolor{brown}{Edge states.--}
One fascinating feature of topological phases is the ability to engineer edge states, which appear at the interface between two distinct topological phases.
Such topological edge states are characterised by near-unity probability at the boundaries, see Fig.~\ref{f0}(b), where the boundary is at the site 0. 
To generate edge states, we have infinitely many options, both using rotation angles as well as $T (\ge2)$ (see Fig.~\ref{f78t2} and SM Sec.~C Figs.~2-8). This yields different winding numbers, e.g., see Fig.~\ref{edge-8} for 8-cycle and also SM Sec. C for 7,4-cycles, with $T=2$. We observe edge states clearly in chaotic (non-periodic) CQWs with 7,8-cycles as periodicity and small cycles can mask the phase boundary effects, see SM Sec. C.

In Fig.~\ref{edge-8}, we consider step-dependent CQW ($T=2$, Fig.~\ref{f78t2}) with a 8-cycle in which position site 0 is acted on by coin ($\theta=\frac{7\pi}{5}$, $\omega= -1$) while other sites are acted on by coin ($\theta=\frac{\pi}{3}$, $\omega= +1$). This defines a boundary at site 0, see Fig.~\ref{f0}(b). We consider the initial state quantum walker,  $\ket{\psi(0)}=\ket{0}_p\otimes\frac{\ket{0}_c+\ket{1}_c}{\sqrt{2}}$. Significant values of probability at site 0 due to the overlap of the walker's initial site with the boundary are characteristic of an edge state. Methods using quantum walks with split-step and split-coin operators (resource-consuming) to create edge states on a 1D line have been shown in Refs.~\cite{kita-pra,pxue,chandra-topo, barry-pra}. We observe clearly long-lived edge states (persistent over time $t$) for 8-cycle in Fig.~\ref{edge-8}, and for 7,4-cycles, see SM Sec.~C.

 We numerically show that these topological edge states in finite cyclic graphs are resilient against small to moderate static and dynamic disorder and also robust against phase (or winding) preserving perturbations, in SM Sec.~D, making them powerful candidates for noise-resilient QI processing and TQC. Further, in  SM Sec.~E, we demonstrate that edge-state generation via our scheme is state independent, i.e., irrespective of the use of a superposed, uneven superposed or unsuperposed initial state. We put forth the algorithms to realize the edge states in cyclic graphs and their resilience against disorders, in SM Sec.~G.
%We can engineer numerous/infinite varieties of edge states by creating a boundary between two distinct phases using Fig.~\ref{f78t2} and SM Sec.~C Figs.~\ref{f78t1}-\ref{f78t5}, \ref{ft3}-\ref{ft4}, in all odd or even cycles, and for other $T>2$ values a similar approach follows too.

\textcolor{brown}{Analysis.--}
We have analytically and numerically proven that both step-dependent (SD, $T\ge 2$) and step-independent (SI, $T=1$) CQWs show topological effects, e.g., topological phases, phase transitions, gapped topological flat bands, Dirac cones for odd-even cycles and rotational flat bands for $4n$-cycles ($n\in \mathbf{N}$). Table~\ref{t1} juxtaposes the key results on both SD and SI-CQW systems with odd and even cycles. In SI-CQW, we do not observe band-closing beyond $k\ne 0$ in odd cycles, unlike even cycles, and this limitation is absent in SD-CQW. A single coin ($\theta=\pi$) yields flat bands in SI-CQW, while multiple coins ($\theta=(2n+1)\frac{\pi}{T}, \;n\in \mathbb{Z_+}\cup \{0\}$) yield flat bands in SD-CQW, for both odd and even cycles. However, rotational flat bands are possible only for even $4n$-cycles ($n\in \mathbf{N}$) in both SD and SI-CQWs. The number of Dirac cones (gap-closing) locations increases with $T$ ($\ge 2$), and SD-CQW shows a larger number of distinct topological phases with topological phase transitions. On the other hand, the SI-CQW does not show any phase transition.
\begin{table}[h!]
    \centering
    \scriptsize
    \resizebox{\columnwidth}{!}{%
    \renewcommand{\arraystretch}{1.5}
    \begin{tabular}{|c|c|c|}
        \hline        
        \textbf{Feature} & \textbf{Odd (3,7) Cycles} & \textbf{Even (4,8) Cycles} \\
        \hline
        \rowcolor{gray!15}\multicolumn{3}{|c|}{\textbf{Step-independent CQW (SI-CQW, $T=1$)}} \\ \hline
        Band closing & \makecell{Yes \& not observed\\ for $k \ne 0$.} & \makecell{Yes, at more locations\\ than odd cycle \& \\observed for $k \ne 0$ too.}\\ \hline
        Flat band & Yes, for one coin. & Yes, for one coin. \\ \hline
        Rotational flat band & No. & Yes. \\ \hline
       \makecell{Topological \\winding number} & \makecell{Topological \\(single value).} & \makecell{Topological \\(single value).} \\ \hline
        Edge states & Not possible. & Not possible. \\ \hline
        \rowcolor{gray!15}\multicolumn{3}{|c|}{\textbf{Step-dependent CQW (SD-CQW, $T \ge 2$)}} \\ \hline
        Band closing & \makecell{No. of locations \\increases with $T$ \& not \\observed for $k \ne 0$ .}& \makecell{No. of locations increases\\ with $T$, more than odd \\cycle \& observed at $k \ne 0$. }\\ \hline
        Flat band & \makecell{Yes, gapped in $k$ and\\ for two or more coins.} & \makecell{Yes, gapped in $k$ and\\ for two or more coins.} \\ \hline
        Rotational flat band & No. & Yes. \\ \hline
        \makecell{Topological \\winding number} & \makecell{Topological \\(multiple values).} & \makecell{Topological \\(multiple values).} \\ \hline
        Edge states & Yes$^{\dagger}.$ & Yes$^{\dagger}.$ \\
        \hline        
    \end{tabular}
    }
    \caption{Comparison of topological features of step-dependent and step-independent CQWs for different cyclic graphs. {\scriptsize{$^{\dagger}$Edge states are long-lived and robust against static and dynamic coin disorder, and phase-preserving perturbations}}.}
    \label{t1}
\end{table}
Moreover, designing topological edge states is possible only with the step-dependent CQW, as it enables to create a phase boundary between two distinct matter phases, unlike the SI-CQW that yields a single topological phase. \noindent Through this study, we show one achieves excellent controllability over the topological effects via CQW on finite cyclic graphs.

For the first time, we obviate the need to use resource-consuming split-step (SS-QW) or split-coin (SC-QW) quantum walks to create edge states, and we use resource-saving small cyclic graphs (periodic lattices). This proposed SD-CQW scheme significantly reduces experimental resources, halving the operator count and requiring a constant number of particle detectors, compared to the $O(2\tau)$ scaling in SS-QW and SC-QW protocols, thereby enhancing the feasibility of topological phase and edge-state generation in photonic and other quantum platforms (see, SM Sec.~F).  We demonstrate with an example time-step of $\tau=100$, that our approach reduces the total experimental resource requirement by a factor of \textbf{$3$}; highlighting its enhanced efficiency and practical implementability for generating topological phase boundaries and edge states.
 %and see SM Sec.~E for a detailed analysis.

%Table~\ref{t1} juxtaposes the features and results on both step-dependent and step-independent variants of CQW and SS-CQW systems for odd and even cycles. See SM Sec.~E and  SM Table~I for a more detailed analysis on our results.

%We also outline a possible experimental implementation of our CQW based scheme in a photonic platform. Therein, single photons act as the quantum walkers; the shift and coin operations are realized entirely with passive optical elements such as beam displacers, polarizing‑beam splitters, waveplates and Jones plates. The quantum walker’s coin state is encoded in photonic polarization and its position sites are encoded in spatial position or orbital-angular-momentum of photons. Site‑specific rotation angles that create the topological phase boundary in real position space can be tuned locally with appropriately oriented half‑ or quarter‑wave plates. After $t$ time-steps, the probability distribution is read out with single‑photon detectors; a pronounced peak at the boundary site will provide the direct experimental signature of the topological edge state created through our scheme.

\textcolor{brown}{Photon-based experimental realization.--} Our scheme can be implemented experimentally using single-photons as quantum walkers, where passive optical elements (waveplates, polarizing-beam-splitters, Jones plates, etc.) mimic the shift \& coin operators. The walker's coin state is encoded in photonic polarization, while the position sites are encoded in spatial modes or orbital-angular-momentum of photons~\cite{bian,pxue,exp1,exp2,exp3,exp4}. Finally, site‑specific rotation angles that create the topological phase boundary can be tweaked locally by appropriately orienting wave/Jones plates. The resulting probability distribution is read out using single‑photon detectors; a pronounced peak at boundary sites will provide the direct experimental signature of edge states created through this scheme.

\textcolor{brown}{Conclusions.--} 
In this work, we introduce cyclic quantum walk (CQW) dynamics on finite cyclic graphs using discrete Fourier transforms and effective Hamiltonian, as a versatile platform for simulating exotic topological phenomena. We demonstrate that both step-independent and step-dependent CQWs offer flexible and resource-saving platforms to generate topological phases (nonzero winding numbers), Dirac cones, topological gapped flat bands and protected edge states. These effects are tunable via step-dependency, site number, periodic evolution and coin-rotation angles. Odd and even cyclic graphs show distinct energy dispersion features, with rotational flat bands emerging exclusively in even $4n$-cycles ($n\in\mathbf{N}$). We derived analytical conditions for the emergence of topological gapped flat bands, confirmed through vanishing group velocity and ill-defined effective mass, and established a direct correspondence between energy gap closings in rotation space and momentum space (Dirac cone). 

Further, we show how to generate topological edge states at the interface between distinct topological phases in both odd and even cycle graphs of finite size. \textbf{Our approach circumvents the need for resource-consuming split-step or split-coin quantum walks to generate edge states}. 
%We emphasize that our approach does not need split-step or split-coin quantum walks, both of which are resource-consuming, in order to generate edge states
We demonstrate that these edge states are robust against static-dynamic disorder and phase-preserving perturbations, and our scheme is independent of the initial states of the quantum walker. This makes the topological phases \& their protected edge states via our scheme highly useful for noise-resilient QI processing and TQC. 
%On topological quantum memory and edge modes: Michnicki, K. P. “3D Topological Quantum Memory with a Power-Law Energy Barrier.” Physical Review Letters 113, 130501 (2014). Physical Review On implementing or simulating edge states / flat-band localization in finite systems: Zhou et al., “Observation of flat-band localization and topological edge states in electrical circuit networks.” Phys. Rev. B 107, 035152 (2023). On quantum communication via flat bands: Almeida et al., “Flat-band quantum communication induced by disorder.” arXiv (2023). On synthetic edge-state platforms and photonic/optomechanical networks: Slim et al., “Programmable synthetic magnetism and chiral edge states in nano-optomechanical quantum Hall networks.” Nature Communications (2025). 
Owing to the finite-dimensional Hilbert space of CQWs, our scheme establishes a highly resource-efficient, experimentally feasible route to engineer and probe topological effects in real systems (e.g., photonic platforms, see SM Sec.~F) and contributes a promising foundation for next-generation fault-tolerant quantum technologies.

Further, our demonstration of robust edge-state formation and flat bands in finite cyclic graphs provides an efficient platform for probing topological protection and enabling experimental realization with minimal resource overhead. The ability to engineer and manipulate such localized edge states without complex split-step or split-coin dynamics highlights their immediate utility in quantum memory architectures~\cite{qmemory, qmemory2}, where protected edge modes can store quantum information with enhanced stability, and in quantum communication networks, using localized modes for noise-resilient state transfer~\cite{edge1, edge2, edge4,qmemory2}.

\clearpage
\newpage
%\vspace{10cm}
\onecolumngrid
\section*{\underline{Supplementary Material} for "Quantum walks reveal topological flat bands, robust edge states and topological phase transitions in cyclic graphs"}

\centerline{Dinesh Kumar Panda$^{1,2}$,\; Colin Benjamin$^{1,2}$}

\centerline{$^{1}$School of Physical Sciences, National Institute of Science Education and Research Bhubaneswar, Jatni 752050, India}
\centerline{$^{2}$Homi Bhabha National Institute, Training School Complex, Anushaktinagar, Mumbai
400094, India}

\vspace{1cm}

In Sec.~\textcolor{blue}{A}, we diagonalize the cyclic quantum walk (CQW) evolution operators using the discrete Fourier transform method to obtain energy dispersions and winding numbers for finite cyclic graph systems, illustrated with explicit examples. We also calculate the group velocity and effective mass of the quantum walker (a quantum particle evolving under CQW dynamics). Numerical results for energy dispersion and topological phases (winding numbers) on 3-, 4-, 7-, and 8-site cycles are presented. In Sec.~\textcolor{blue}{B}, we provide rigorous theoretical proofs for the conditions under which topological flat bands and rotational flat bands emerge, as well as the appearance of Dirac cones (linear band closings) in momentum and rotation spaces, and demonstrate their equivalence. Sec.~\textcolor{blue}{C} details the construction of topological edge states on both odd- and even-site cyclic graphs. 
In Sec.~\textcolor{blue}{D}, we prove the robustness of these topological edge states against moderate static and dynamic disorder in gate (coin) operations and establish their resilience to phase-preserving perturbations.
In Sec.~\textcolor{blue}{E}, we demonstrate that our step-dependent CQW (SD-CQW) scheme for generating edge states is state-independent, i.e., it is completely independent of the nature (superposed, out-of-phase superposed, uneven superposed and unsuperposed) of the initial coin state (initial state) of the quantum walker. We compute the resource overhead (operator count plus detector count) of our SD-CQW scheme in Sec.~\textcolor{blue}{F}. Our protocol halves the operator count and keeps the number of particle detectors constant, whereas existing split-step and split-coin schemes require both operators and detectors to scale linearly with the number of time steps ($\tau$), greatly enhancing experimental feasibility. In Sec.~\textcolor{blue}{G}, we present an algorithm along with Python codes for generating edge states in cyclic graphs and to analyze the effects of disorder and different initial states on their stability. Finally, Sec.~\textcolor{blue}{H} concludes with a comprehensive analysis summarizing the results presented in this Supplementary Material (SM).

\subsection{Calculation of energy dispersion and topological phases in cyclic graphs}
\subsubsection{Analytical results on energy dispersion, effective mass and group velocity}
As discussed in the main text page 2, part on \textit{"Model"}, a cyclic quantum walk (CQW) describes the propagation of the spatial distribution of a single quantum particle (e.g., electron or photon)
on an $N$-cycle graph, i.e., on $N$ sites of a cyclic graph (e.g., atomic sites or position or orbital angular momentum). The quantum walker or single quantum particle lives in a composite Hilbert space $\mathcal{H} = \mathcal{H}_P \otimes \mathcal{H}_C $, composed of an $N$-dimensional position space ($\mathcal{H}_P$ spanned by $\{\ket{x}: x \in {0,1,2,\dots,k-1}\}$) and a 2-dimensional coin space ($\mathcal{H}_C$ spanned by $\{\ket{0}_c,\ket{1}_c\}$). The walker's motion can be clockwise or anticlockwise, which is governed by a translation/shift operator $\hat{S}$ contingent upon the action of a single-qubit gate (coin operator) $\hat{C_2}$.
CQW has spatial symmetry and can be diagonalized via Fourier transform methods~\cite{kita-pra,pxue,chandra-topo,asboth,karimi19,filippo17,panahiyan20}, i.e, the spatial computation basis vector $\ket{x}$ can be mapped as, 
\begin{equation}
         |x\rangle = \frac{1}{\sqrt{N}} \sum_{k'=0}^{N-1} e^{-i\frac{2\pi}{N}k'x} |k'\rangle, \;\text{thus, } |k'\rangle = \frac{1}{\sqrt{N}} \sum_{x=0}^{N-1} e^{i\frac{2\pi}{N}k'x} |x\rangle, 
        \label{eq1}
 \end{equation}
where range of $ |k'\rangle $ is same as that of $ |x\rangle $, i.e., $k',x \in [0, N-1]$. Further, $\sum_{k'} |k'\rangle \langle k'| = \mathbb{1}$ with $\bra{k'}\ket{k''}=\delta_{k'k''}$ confirms the completeness of the quasi-momentum basis \{$\ket{k'}$\}  (of the periodic system) . Note, $N$ can be even or odd, i.e., even cycle graphs: 4-cycle, 6-cycle, 8-cycle..., or odd cycle graphs: 3-cycle, 5-cycle, 7-cycle,...

The quantum walker on the cyclic graphs moves anticlockwise (clockwise) by one site for coin state $\ket{0}_c$ ($\ket{1}_c$) and is achieved via a unitary shift/translation operator, 
\begin{equation}
    \hat{S} = \sum_{q=0}^{1}\sum_{x=0}^{N-1}\ket{(x+(-1)^{q}) \text{ mod } N}\bra{x}\otimes\ket{q}_c\bra{q}_c.
    \label{shift}
\end{equation}

The complete time-evolution of such quantum particle is characterized by,
\begin{equation}
U_{N}(t) = \hat{S}.[I_N\otimes \hat{C}_{2}(\theta, T)]\;,
\label{eq2}
\end{equation}
\noindent
and the evolved quantum state at $t$ time step is,
\begin{equation}
    \ket{\psi(t)} = U(t)\ket{\psi(t-1)} = U_{k}(t)U_{k}(t-1)...U_{k}(1) \ket{\psi(0)}.
    \label{stateevol}
\end{equation}
\noindent
We find that the translation operator (non-diagonal in computation basis), Eq.~(\ref{shift}), is diagonal in momentum basis (Eq.~(\ref{eq1})), i.e., 
\begin{equation}
    \hat{S} \ket{k'}\ket{q}_c=\lambda_q\ket{k'}\ket{q}_c \text{ where, } \lambda_q=e^{(-1)^{q+1}\frac{2\pi i}{N} k'}.
\end{equation}
Here, $\lambda_q$'s define the eigenvalues of the translation $\hat{S}$ with $ q\in\{0,1\}$. Thus, the translation operator, applicable for all $k'$ values, takes the form,
\begin{equation}  
\hat{S} = {|k' \rangle \langle k'|} \otimes \sum_{k'=0}^{N-1} 
\begin{pmatrix} 
e^{-i\frac{2\pi}{N}k'} & 0 \\ 
0 & e^{i\frac{2\pi}{N}k'} 
\end{pmatrix}  \equiv e^{-i\frac{2\pi}{N}k'\sigma_z}
 \label{eq3}
\end{equation}
 The 2D or single-qubit gate (coin operator) has the general form,  
 $\hat{C}_{2}(\theta, T)\equiv e^{-i \frac{T\theta}{2} \sigma_y}$ where the rotation angle $\theta \in [0,2\pi]$. Here, $T=1$ denotes a step-independent coin operation, and the step-independent arbitrary coin has the form,
 \begin{equation}
      \hat{C}_{2}(\theta, T=1)= e^{-i \frac{\theta}{2} \sigma_y}=
 \begin{pmatrix} 
\cos{\frac{\theta}{2}} & -\sin{\frac{\theta}{2}} \\ 
\sin{\frac{\theta}{2}} & \cos{\frac{\theta}{2}}
\end{pmatrix}.
 \label{eq4}
 \end{equation}
For the special case $\theta=\frac{\pi}{2}, T=1$, $\hat{C}_{2}(\frac{\pi}{2}, 1)$ gives the Hadamard gate/coin.
We analyze both step-dependent ($T>1$) and step-independent ($T=1$) CQW evolution. Due to the unitarity of the CQW evolution,  we can transform the unitary evolution to a stroboscopic evolution~~\cite{kita-pra,pxue,chandra-topo,asboth} via an effective Hamiltonian $\hat{H}$ (in units of $\hbar=1$),
\begin{equation}
\hat{H} = E(k) \hat{n}(k) \cdot \vec{\sigma}, \text{ with, } U = e^{-i H}.\;  
\label{hamil}
\end{equation}

Here, $ E(k)$ denotes the energy dispersion, $\vec{\sigma}$ consists of the Pauli matrices and $\hat{n}(k)$ refers to the eigenstates of the quantum walker (particle).

Plugging the expression of the coin and translation operators in $k'$-space, i.e., $\hat{C}_{2}(\theta, T)\equiv e^{-i \frac{T\theta}{2} \sigma_y}$ and $\hat{S}=e^{-i\frac{2\pi}{N}k'\sigma_z}$ in Eq.~(\ref{hamil}) for $U = e^{-i E(k) \hat{n}(k) \cdot \vec{\sigma}}$, we get the energy dispersion relation for an arbitrary $N$-site cyclic graph as,
\begin{equation}
\begin{split}
     E(k) = \pm \arccos(\cos k \cos \frac{T\theta}{2}), 
    \text{ or, }
    E(k') = \pm \arccos(\cos \frac{2\pi k'}{N} \cos \frac{T\theta}{2}),
    \label{eq6}
    \end{split}
\end{equation}

and the winding vector $\vec{n}$,
\begin{equation}
\hat{n} = \begin{pmatrix}
n_x \\ n_y \\n_z
\end{pmatrix}=\frac{1}{\sin E(k')} 
\begin{pmatrix}
- \sin \frac{2\pi k'}{N} \sin \frac{T\Theta}{2} \\
\cos \frac{2\pi k'}{N} \sin \frac{T\Theta}{2} \\
\sin \frac{2\pi k'}{N} \cos \frac{T\Theta}{2}
\end{pmatrix}.
 \label{eq7}
\end{equation}
The two energy bands (upper and lower bands) from Eq.~(\ref{eq6}) correspond to the two internal states of the quantum particle/walker. The associated Hamiltonian, $\hat{H}$ in Eq.~(\ref{hamil}), are traceless, and the energy bands have the symmetry, $E(k) = E(-k)$ or equivalently $E(k') = E(-k')$, spanning $[-\pi, \pi]$ with $k\in[0,2\pi]$, and $k=\frac{2\pi k'}{N}$ with $k'\in [0, N-1]$. 
The energy band gap closures occur at $E = 0$ and $\pm \pi$. 
We observe only Dirac cones (see Fig.~\ref{f0}(a)) featuring energy bands that are linear in momentum $k$, leading to a gap closure in the CQW evolution, see subsection 3 (Theorem 1). 

%Topological insulating phases can emerge when these bands are gapped, while edge states arise at the interface of these phases. 

We define $k=\frac{2\pi k'}{N}$ such that  $ k' \in [0, N - 1]$ and \( k \) runs over a complete cycle of values in \( [0, 2\pi] \).  
As the number of nodes grows, the discrete \( k \) values gradually form a continuum~\cite{karimi19}. Thus, the dispersion relation becomes,
\begin{equation}
    E(k) = \pm \arccos(\cos k \cos \frac{T\theta}{2}).
     \label{eq8}
\end{equation}
 Clearly, for step-dependent CQW dynamics ($T>1$), one can observe a large variety of dispersion and band closing (e.g., more number of Dirac cones) and hence rich topological features (multiple topological invariants can be controllably obtained for different locations in coin parameter space with respect to $T$), as compared to step-independent CQW ($T=1$), which we see graphically too, in main text Figs. 2 and 3 for finite $N$ site cyclic graphs.
The group velocity for the walker/ quantum particle can be calculated as,
\begin{equation}
v_{gr}(k,\theta,T) = \frac{\partial E(k)}{\partial k} = \pm \frac{ \cos\frac{T\theta}{2} \sin k}{\sqrt{1 - \cos^2 \frac{T\theta}{2} \cos^2 k}}.
\label{eq9}
\end{equation}
%which is shown in Supplementary Fig.~\ref{fs2} vs $k$ and $\theta$ for different $T$ values, and the result agrees with the energy dispersion slopes where we observe Dirac cones.
Clearly, for flat bands, the group velocity $\rightarrow0$, the conditions (i.e., values of $\theta, T, N$) of which are derived in subsection~3 (Theorem~2). 
The quantum particle's effective mass, which arises due to the curvature of energy bands, takes the form\cite{effmass1,effmass2,effmass3},
\begin{equation}
m^{*}(k,\theta,T)= \frac{\hbar^2}{\frac{\partial^2 E(k)}{\partial k^2}} = \frac{1}{\frac{\partial v_{gr}}{\partial k}}= \pm\frac{(1 - \cos^2\frac{T\theta}{2} \cos^2 k)^{3/2}}{\cos k \cos\frac{T\theta}{2} \sin^2\frac{T\theta}{2}},
 \label{eq10}
\end{equation}
\text{(in units of } $\hbar = 1$\text{)}, where the + (-) sign corresponds to the upper (lower) energy band. For flat bands (topological),  $m^{*}(k,\theta,T)\rightarrow \infty$ and this can happen when the group velocity $\rightarrow0$. Thus, groups velocity and effective mass can work as a measure/signature of flat band occurrences in the quantum system. 
%In solid state matter, effective mass plays a key role in determining carrier mobility, diffusion rate, and spreading in wave packets~\cite{effmass1,effmass2,effmass3}. Low effective mass means fast spreading, a lighter inertia and higher mobility. Understanding and determining the effective mass is thus pivotal for the advancement of modern electronics, optoelectronics, and quantum technologies.
In solid-state systems, the effective mass plays a key role in determining carrier mobility, diffusion rates, and wave packet spreading~\cite{effmass1,effmass2,effmass3}. A low effective mass corresponds to faster spreading, lower inertia, and higher mobility. Accurately understanding and determining the effective mass is therefore pivotal for advancing modern electronics, optoelectronics, and quantum technologies. As shown in Eq.~\ref{eq10}, the effective mass can be controlled via two parameters $\theta$ and $T$.  
\begin{figure}[h]
\includegraphics[width = 13cm,height=4.5cm]{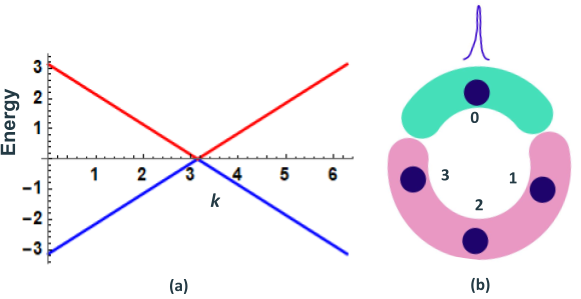}
\caption{Schematics of (a) a Dirac cone for CQW, where energy gap closing is linear; (b) two distinct topological phase regimes are shown in green and red on a 4-cycle and an edge state (wave peak in blue) is expected to form at the boundary between the phases.}
% and a nonlinear gap closing refers to Fermi arc
\label{f0}
\end{figure}
\subsubsection{Analytical results for topological phases and winding numbers}We can derive the relationship for the Berry phase, associated with the state resulting from the cyclic Hamiltonian under adiabatic evolution. Given that this system is one-dimensional and periodic, the Berry phase is referred to as the Zak phase, for derivation of this refer to Refs.~\cite{karimi19,filippo17,panahiyan20}, and it is calculated via, 

\begin{equation} Z_{\theta,T} = \frac{1}{2} \oint dk \left( \hat{n}(k) \times \frac{\partial \hat{n}(k)}{\partial k} \right) \cdot \hat{A}(\theta,T). \label{eq11}  \end{equation}
Here, $\hat{A}(\theta,T)$ is unit vector in the Bloch sphere which is perpendicular to $\hat{n}, \forall k$, and $\hat{n}$ is derived in Eq.~\ref{eq7}. One can find using Eq.~(\ref{eq7}), $\hat{A}(\theta,T)=\left( \cos\frac{T\theta}{2}, 0, \sin\frac{T\theta}{2} \right)$. The chosen unit vector $\hat{A}(\theta,T)$ corresponds to a direction that is constant in $k$ or $ k'$, which it must, to respect gauge-invariance and to avoid picking extraneous terms through the integration over $k$ or summation over $k'$~\cite{gauge, karimi19}.
The closed integral spans the full Brillouin zone, i.e., $k\in[0,2\pi]$. With some algebraic manipulations, the Zak phase can be simplified for any rotation angle $\theta$ and time-dependency $T$, as follows:

\begin{equation}
\begin{split}
    Z_{\theta,T} = -\frac{\sin(\frac{T\theta}{2})}{2} \oint \frac{dk}{\cos^{2}(\frac{T\theta}{2})\cos^{2}(k) - 1}. 
\end{split}
\label{eq12} 
\end{equation}

For CQW with the Hadamard coin with $\theta=\frac{\pi}{2}$, which creates equal superposition in the coin basis and with step-independency case ($T=1$), the Zak phase is, $Z_{\frac{\pi}{2},1} =\pi$~\cite{karimi19}, i.e., the winding number is 1.

For a cyclic graph or a discretized system with $N$ nodes, using an arbitrary coin $\hat{C}_{2}(\theta, T)$, we find Zak phase and winding number $\omega_{\theta,T,N}=\frac{Z_{\theta,T,N}}{\pi},$ as follows:
\begin{equation}
  Z_{\theta,T,N}=
\sum_{k'=0}^{N-1}\frac{\pi \sin[\frac{T\theta}{2}]}{N(1-\cos\left[\frac{2\pi k'}{N}\right]^2 \cos[\frac{T\theta}{2}]^2)}, \quad \omega_{\theta,T,N}=
\sum_{k'=0}^{N-1}\frac{ \sin[\frac{T\theta}{2}]}{N(1-\cos\left[\frac{2\pi k'}{N}\right]^2 \cos[\frac{T\theta}{2}]^2)}.
\label{eq13}
\end{equation}
In particular, for the Hadamard coin and without step-dependency ($T=1$), the winding number reduces to, $
\sum_{k'=0}^{N-1}\frac{-2 \sqrt{2}}{N \left( -3 + \cos \left[ \frac{4 k' \pi}{N} \right] \right)},
 $
which for large $N$ approximates to the value derived above, i.e., $\omega_{\frac{\pi}{2},1} =1$, and for $N=5$, (i.e., a 5-cycle CQW), we get, $\omega_{\frac{\pi}{2},1,5} \approx  \frac{29\sqrt{2}}{41}= 1.0003 \approx 1.$
This value changes slightly with $N$ (not qualitatively), for example with $T=1$ (step-independent CQW), for $N=3$ we get $\omega_{\frac{\pi}{2},1,3} \approx 1.01015$,  for $N=4, \omega_{\frac{\pi}{2},1,4} \approx 1.06066$, for $N=8, \omega_{\frac{\pi}{2},1,8} \approx 1.00173$, for $N=7$ (7-cycle) we get $\omega_{\frac{\pi}{2},1,7} \approx 1.00001$, and when $N\rightarrow$ very large (say $N=1000$), winding number becomes exactly $1$ for the Hadamard coin as we obtained for the continuum $k$ case. Similarly, for $T=2$ (step-dependent CQW) and a coin with rotation angle $\theta=\frac{3\pi}{2}$, we get for $N=3, \omega_{\frac{3\pi}{2},2,3} =-1$,  for $N=4, \omega_{\frac{3\pi}{2},2,4} =-1$, for $N=8, \omega_{\frac{3\pi}{2},2,8} =-1$, which are all equivalent as $\omega_{\frac{\pi}{2},2,1000} =-1$ for $N=1000$. For a separate coin ($\frac{\pi}{3}$), $\omega_{\frac{\pi}{3},2,7}=1$ for 7-cycle and $\omega_{\frac{\pi}{3},2,8} \approx 1.00005$ for 8-cycle, which are equivalent as $, \omega_{\frac{\pi}{3},2,1000}=1$ for large $N=1000$-cycle. Thus, as we see for the $N=3,4,5,7,8,...$ cases above, it is reasonable to study topological effects with these small cycle graphs as these are less resource-intensive and more feasible to generate experimentally due to their smaller working Hilbert space~~\cite{bian, karimi19,pral}, than any large, multidimensional or 1D infinite line or lattices. This will help simulate topological phases, band closing, and edge states in physical systems in a resource-saving manner in experiments, such as with photonic or ion-trap circuits~\cite{bian, karimi19}.

Furthermore, odd and even small cycle graphs with $T>1$ and arbitrary $\theta$ values may yield distinct behaviour in energy dispersion (at different coin parameters and $T$) and topological features, which we discuss in the main text \textit{"Results"} part, and provide some more examples in the following discussions and figures below. 
  
\subsubsection{Numerical results for energy dispersion and topological phases with CQW}

\begin{figure}[H]
\includegraphics[width = 18cm,height=6.2cm]{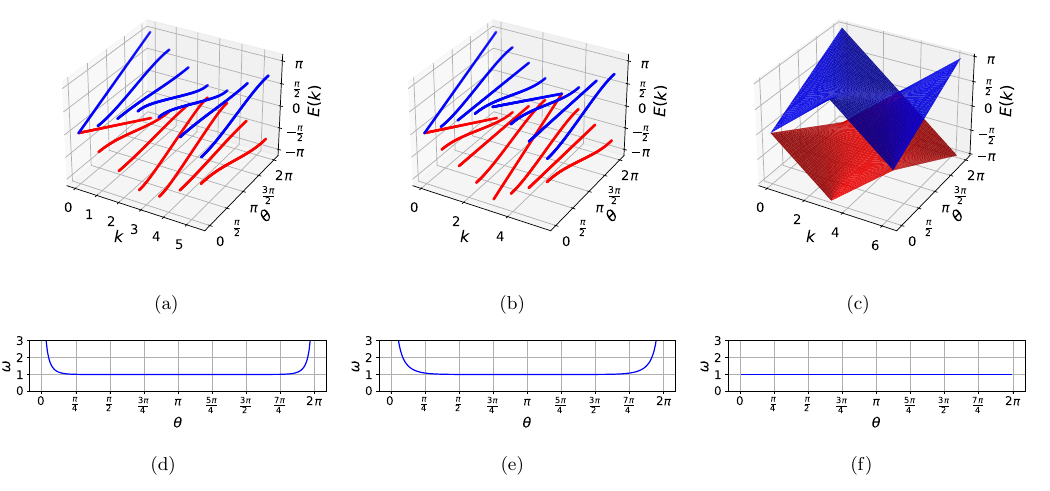}
\caption{Energy dispersion vs quasi-momenta $k$ and rotation angle $\theta$ for (a) $N=7$, (b) $N=8$-cycles and (c) $N=1000$ (with Dirac cones). The blue (red) surface refers to the upper (lower) energy band.; Winding number $\omega$ vs $\theta$ for (d) $N=7$, (e) $N=8$ and (f) $N=1000$ ($k$ continuum limit), for step-independent ($T=1$) CQW. }
\label{f78t1}
\end{figure}

In the main text, we show the energy dispersion and topological invariant: winding number ($\omega$) for 7 and 8 cycles with step-dependent CQW ($T=2$).
Below, Figs.~\ref{f78t1}(a-c) shows the energy dispersion vs quasi-momenta $k$ and rotation angle $\theta$ for $N=7$, $N=8$ and $N=1000$-cycles, for step-independent ($T=1$) CQW. The blue (red) surface refers to the upper (lower) energy band. Further, Figs.~\ref{f78t1}(d-f) shows winding number $\omega$ vs $\theta$ for for $N=7$, $N=8$ and $N=1000$-cycles, for step-independent ($T=1$) CQW.

%\end{widetext}

\begin{figure}[H]
\includegraphics[width = 17cm,height=7cm]{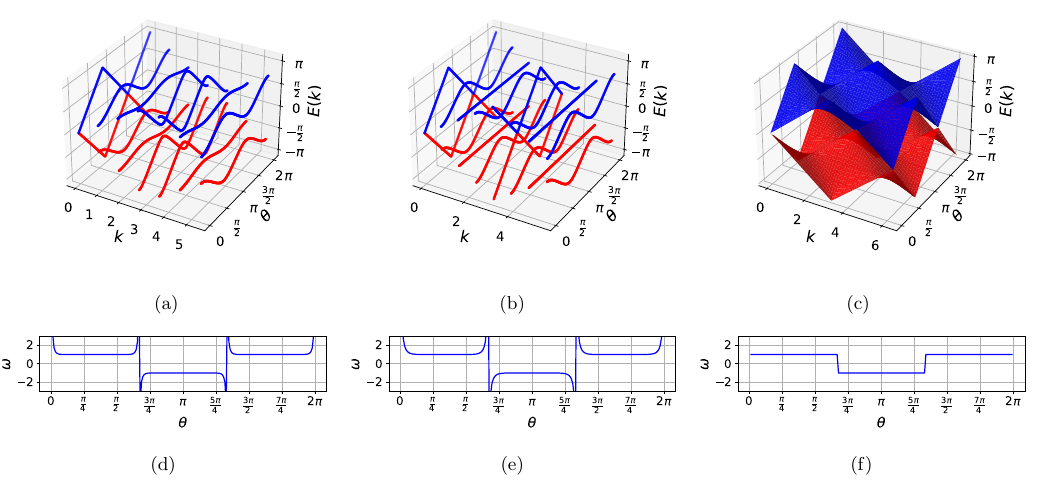}
\caption{Energy dispersion relation vs quasi-momenta $k$ and modified rotation angle $\theta$ for (a) $N=7$, (b) $N=8$-cycles and (c) $N=1000$ (large $N$ limit, with Dirac cones) with step-dependent ($T=3$) CQW. The blue surface (band) is related to the upper energy band, and the red surface is associated with the lower energy band. Band closing happens at $E=0,\pm \pi$ with Dirac cones. Winding number $\omega$ vs rotation angle $\theta$ for (e) $N=7$, (f) $N=8$ and (g) $N=1000$ ($k$ continuum limit), for  $T=3$ (step-dependent CQW). }
\label{f78t3}
\end{figure}

\begin{figure}[H]
\includegraphics[width = 17cm,height=7cm]{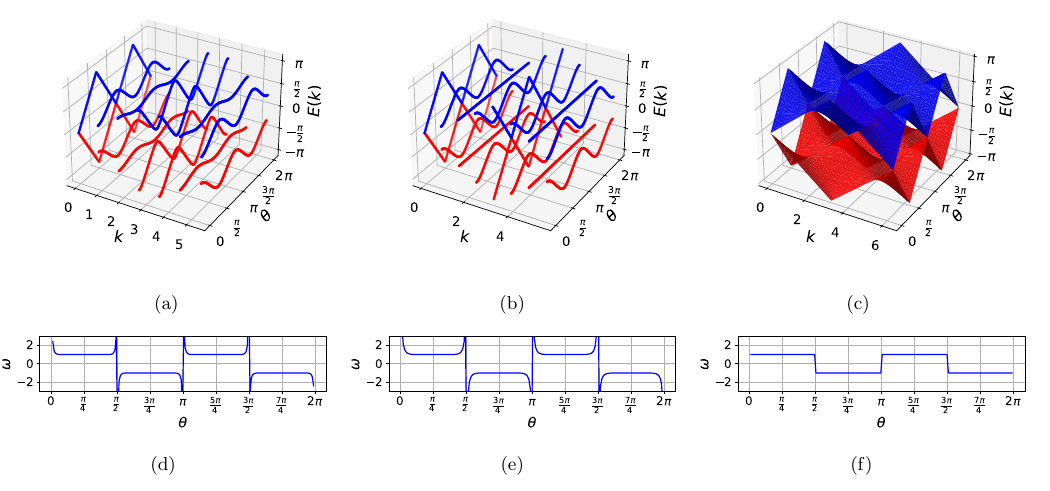}
\caption{Energy dispersion relation vs quasi-momenta $k$ and modified rotation angle $\theta$ for (a) $N=7$, (b) $N=8$-cycles and (c) $N=1000$ (large $N$ limit, with Dirac cones) with step-dependent ($T=4$) CQW. The blue surface (band) is related to the upper energy band, and the red surface is associated with the lower energy band. Band closing happens at $E=0,\pm \pi$ with Dirac cones. Winding number $\omega$ vs rotation angle $\theta$ for (e) $N=7$, (f) $N=8$ and (g) $N=1000$ ($k$ continuum limit), for  $T=4$ (step-dependent CQW). }
\label{f78t4}
\end{figure}

\begin{figure}[H]
\includegraphics[width = 17cm,height=7cm]{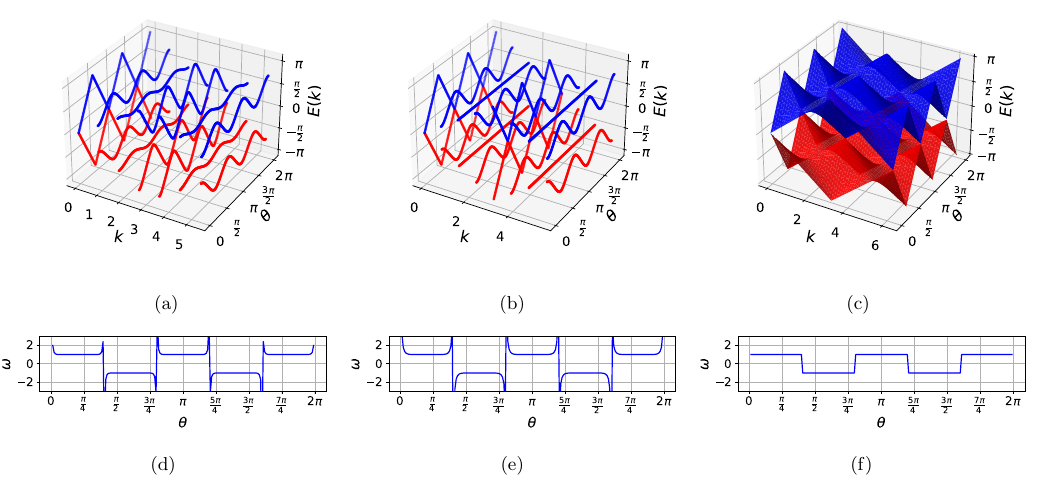}
\caption{Energy dispersion relation vs quasi-momenta $k$ and modified rotation angle $\theta$ for (a) $N=7$, (b) $N=8$-cycles and (c) $N=1000$ (large $N$ limit, with Dirac cones) with step-dependent ($T=5$) CQW. The blue surface (band) is related to the upper energy band, and the red surface is associated with the lower energy band. Band closing happens at $E=0,\pm \pi$ with Dirac cones. Winding number $\omega$ vs rotation angle $\theta$ for (e) $N=7$, (f) $N=8$ and (g) $N=1000$ ($k$ continuum limit), for  $T=5$ (step-dependent CQW). }
\label{f78t5}
\end{figure}

Further, Figs.~\ref{f78t3}-\ref{f78t5} show the energy dispersion and winding number ($\omega$) for step-dependent CQW with $T=3,4,5$ for 7 and 8-cycles (i.e., cyclic graphs with 7 and 8 sites), respectively.

\begin{figure}[H]
\includegraphics[width = 17cm,height=7cm]{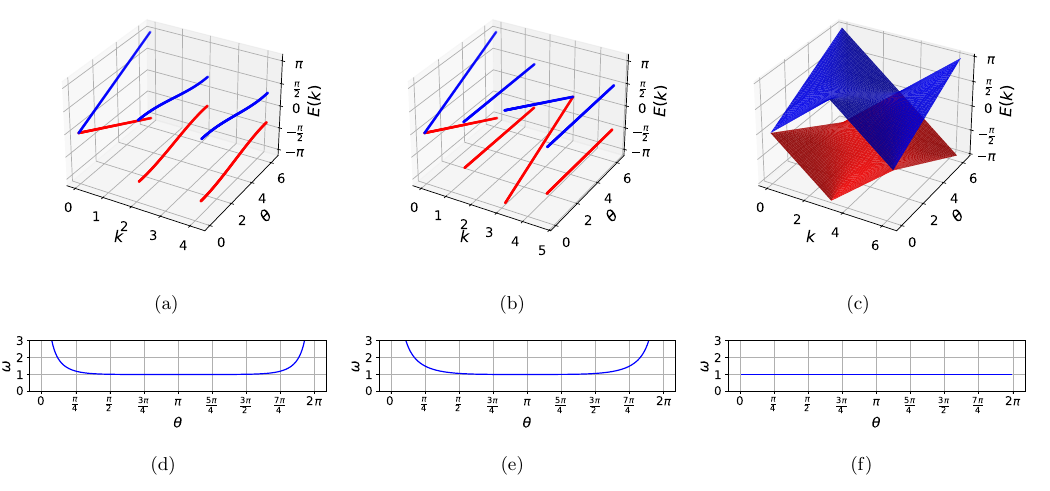}
\caption{ Energy dispersion relation vs quasi-momenta $k$ and modified rotation angle $\theta$ for (a) $N=3$, (b) $N=4$-cycles and (c) $N=1000$ (large $N$ limit, with Dirac cones) with step-independent ($T=1$) CQW. The blue surface (band) is related to the upper energy band, and the red surface is associated with the lower energy band. Band closing happens at $E=0,\pm \pi$ with Dirac cones. Winding number $\omega$ vs rotation angle $\theta$ for (e) $N=3$, (f) $N=4$ and (g) $N=1000$ ($k$ continuum limit), for  $T=1$ (step-independent CQW). }
\label{ft1}
\end{figure}

\begin{figure}[H]
\includegraphics[width = 17cm,height=7cm]{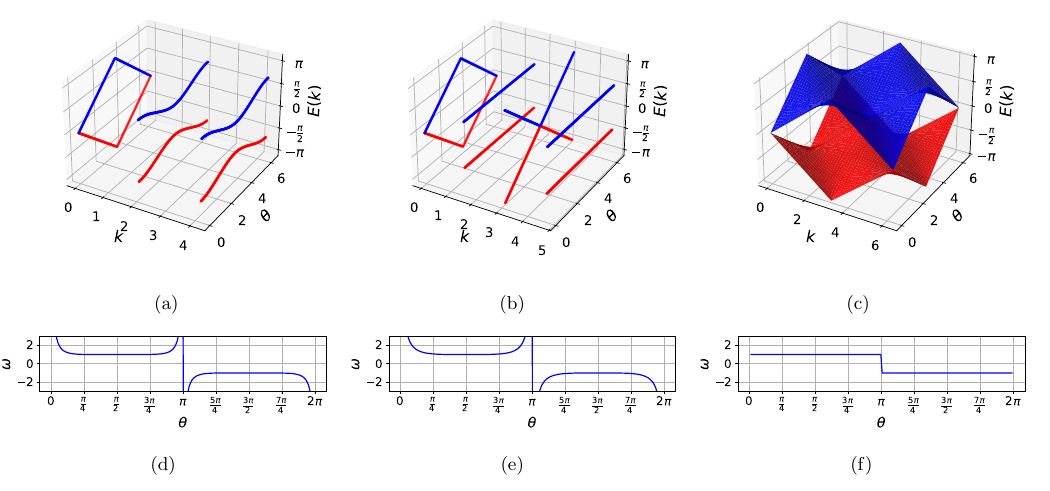}
\caption{Energy dispersion relation vs quasi-momenta $k$ and rotation angle $\theta$ for (a) $N=3$, (b) $N=4$-cycles and (c) $N=1000$ (large $N$ limit, with Dirac cones shown), with step-dependent ($T=2$) CQW. The blue surface is related to the upper energy band, and the red surface is associated with the lower energy band. Band closing happens at $E=0,\pm \pi$ with Dirac cones. Winding number $\omega$ vs rotation angle $\theta$ for (e) $N=3$, (f) $N=4$ and (g) $N=1000$ ($k$ continuum limit), for  $T=2$ (step-dependent CQW). }
\label{ft2}
\end{figure}

\begin{figure}[H]
\includegraphics[width = 17cm,height=7cm]{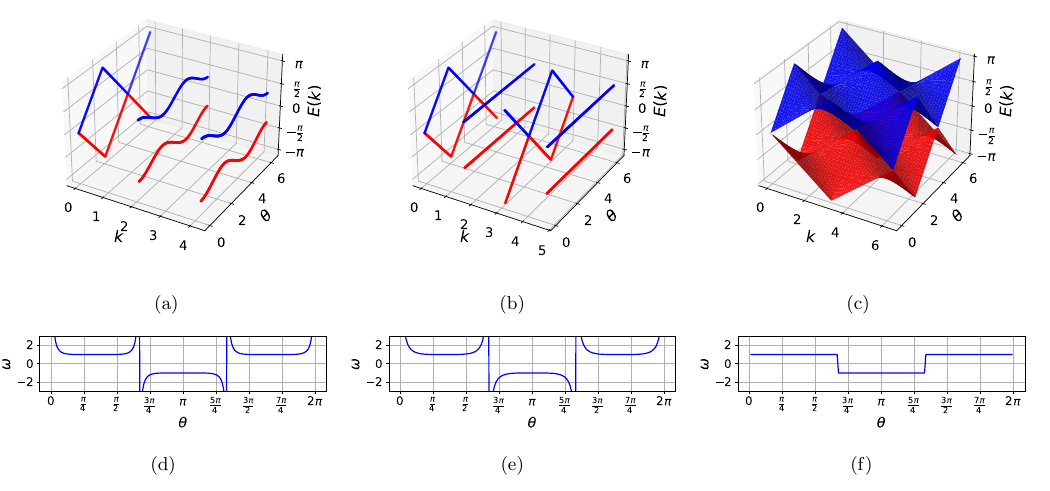}
\caption{Energy dispersion relation vs quasi-momenta $k$ and rotation angle $\theta$ for (a) $N=3$, (b) $N=4$-cycles and (c) $N=1000$ (large $N$ limit, with Dirac cones shown), with step-dependent coin ($T=3$) or CQW. The blue surface is related to the upper energy band, and the red surface is associated with the lower energy band. Band closing happens at $E=0,\pm \pi$ with Dirac cones. Winding number $\omega$ vs rotation angle $\theta$ for (e) $N=3$, (f) $N=4$ and (g) $N=1000$ ($k$ continuum limit), for  $T=3$ (step-dependent CQW). }
\label{ft3}
\end{figure}

\begin{figure}[H]
\includegraphics[width = 17cm,height=7cm]{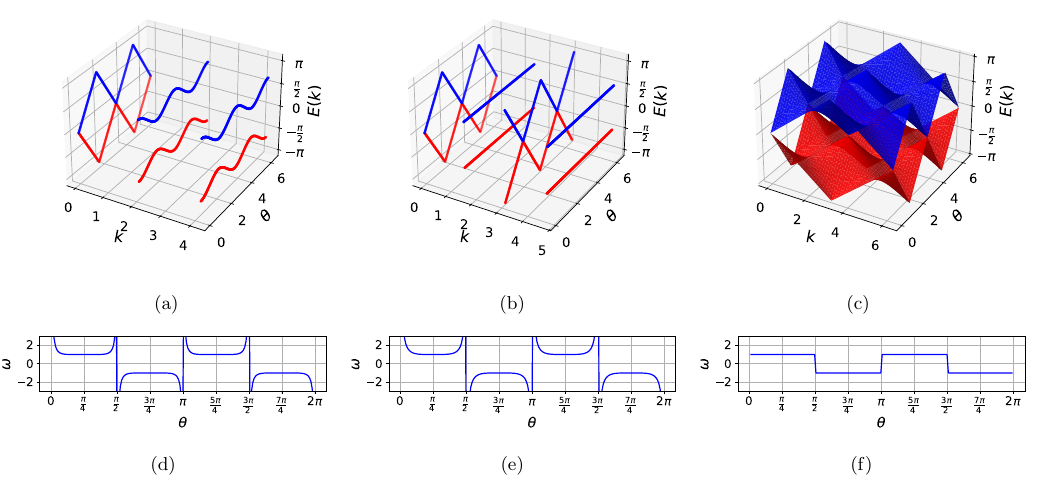}
\caption{Energy dispersion relation vs quasi-momenta $k$ and rotation angle $\theta$ for (a) $N=3$, (b) $N=4$-cycles and (c) $N=1000$ (large $N$ limit, with Dirac cones shown), with step-dependent coin ($T=4$) or CQW. The blue surface is related to the upper energy band, and the red surface is associated with the lower energy band. Band closing happens at $E=0,\pm \pi$ with Dirac cones. Winding number $\omega$ vs rotation angle $\theta$ for (e) $N=3$, (f) $N=4$ and (g) $N=1000$ ($k$ continuum limit), for  $T=4$ (step-dependent CQW). }
\label{ft4}
\end{figure}

Further, the energy dispersion and topological invariant: winding number ($\omega$) for 3 and 4-cycles (i.e., cyclic graphs with 3 and 4 sites)  with $T=1,2,3,4$, are shown in Figs.~~\ref{ft1}-\ref{ft4} respectively. 

%\begin{widetext}  

  
Notably, the trend in winding number with respect to rotation $\theta$ for both odd and even cycles is identical and odd-even cycle distinction vanishes as $N$ becomes large, e.g., $N=1000, 1001...$ However, for small finite cycles, like 3, 4, 7, 8-cycles, the odd-even distinction is relevant in the energy dispersion and band-closing and flat bands, see main text \textit{Results} page 3. 

We observe that with increasing $T$, the number of locations where the energy band-gap closes increases, and so does the number of edge states for both 3-, 4- and very large ($N=1000$)-cycles. The increase in the number of edge states is evident from the increased variation of nonzero winding numbers, with increasing $T$, see Fig.~\ref{ft2}-\ref{ft4} in comparison to Fig.~\ref{ft1}. Herein, a nonzero (zero) winding number indicates a topological (trivial) phase of the quantum systems evolving via CQW dynamics, and for step-dependent coins ($T>1$ cases) shows a larger number of distinct topological phases (i.e., more number of different winding numbers) than the step-independent coins ($T=1$ case).

One distinct feature we see in the 4-cycle (see Figs.~\ref{ft1}-\ref{ft2} (b)) as compared to the 3-cycle (see Figs.~\ref{ft1}-\ref{ft2} (a)) is that the number of band-closing locations is larger in the 4-cycle. Besides, we see band closing beyond trivial $k=0$, such as at $k=\pi$ only in the 4-cycle case for particular coins (rotation angle $\theta$), and it holds for both step-independent and step-dependent CQW, i.e., for any $T$ values, see Figs.~\ref{ft1}-\ref{ft4}.

\subsection{Theorems and Proofs: Dirac cones and Flat bands in rotation and momentum spaces}
In the main text \textit{Results} part, we have mentioned about energy-gap closing in rotation space implies energy gap-closing (Dirac cones) in momentum space, a condition for the generation of flat bands (topological) and occurrence of rotational flat bands in $4n$-cyclic graphs ($n\in \mathbf{N}$). Herein, we prove these facts with analytical derivations taking recourse to the generalized energy dispersion given in Eq.~(\ref{eq6}) and generalized group velocity given in Eq.~(\ref{eq9}) for CQW evolution, as follows. 
\\

\textbf{Theorem 1a: }
Energy gap closing in rotation space implies energy gap-closing (Dirac cones) in momentum space.

\textbf{\textit{Proof:}} As shown in Figs.~\ref{f78t3}-\ref{f78t5} (a-c) for 7,8-cycles and Figs.~\ref{ft1}-\ref{ft4} (a-c) for 3,4-cycles, energy gap closing in coin parameter $\theta$ implies Dirac cones (linear gap closing in $k$). One can analytically show this, using energy dispersion as in Eq.~(\ref{eq6}), the energy band-gap closing at $E(k)=0$ implies,

\begin{equation}
\begin{split}
   \cos^{-1}(&\cos k  \cos \frac{T\theta}{2})=0
   \;\implies \cos k \cdot \cos \frac{T\theta}{2}= 1 
 \text{ or, } \cos k\; , \cos \frac{T\theta}{2}=\pm 1,
   \\& \text{i.e., }
   \theta = 
\begin{cases}
0,\frac{4 n \pi}{T}; &  k=0, 2\pi \;(\text{or, } k' = 0, N) \\
\frac{(4n+2) \pi}{T}; & k=\pi\;(\text{or, }k' = \frac{N}{2})
\end{cases}, \text{ where } n\in\{0,1,2,3,...\} .
\end{split}
\label{eq31}
\end{equation}
From the condition in Eq.~(\ref{eq31}), we see that one can control the energy gap closing and Dirac cone location by tuning the CQW parameters $T, \theta, N$.

Rotation $\theta$ values shown in Eq.~$\ref{eq31}$ refer to both lower and upper band energy to become $E(k)=0$, i.e., energy gap closing, at momentum values $k=0,\pi,2\pi$. This means the upper and lower energy bands will close their gap in $k$ for those $\theta$ values in Eq.~$\ref{eq31}$. Thus, an energy gap closing in coin parameter $\theta$ space implies energy gap-closing (Dirac cones) in momentum $k$ (or, $k'$) space, too.

For instance, in Figs.~\ref{ft1}(a-b) for $T=1$, at $\{k=0, \theta=0\}$ in a 3-cycle, then at $\{k=0, \theta=0\}$ and $\{k=\pi, \theta=2\pi\}$ in a 4-cycle, we see energy gap closing in rotation $\theta$ space. These $\theta$ values show Dirac cones (gap-closing) in $k$ ($N\rightarrow \infty$, continuum limit) too, see Fig.~\ref{ft1}(c). Notably, an odd-cycle graph (e.g., 3-cycle, 7-cycle) does not show energy band-closing at $k\ne 0,2\pi$, unlike even-cycle graphs (e.g., 4-cycle, 8-cycle). 

Moreover, from Eq.~(\ref{eq31}) and Fig.~\ref{ft2}, we see Dirac cones at $\theta=0,\pi, 2\pi$ for $T=2$, and the number of gap closing points increases with $T$, see Fig.~\ref{ft2}-\ref{ft4}. Similar results are also observed in 7 and 8-cycles, too, see  Figs.~\ref{f78t3}-\ref{f78t5} and Figs.~2-3 in the main.
\\

\begin{figure}[H]
\includegraphics[width = 18cm,height=9cm]{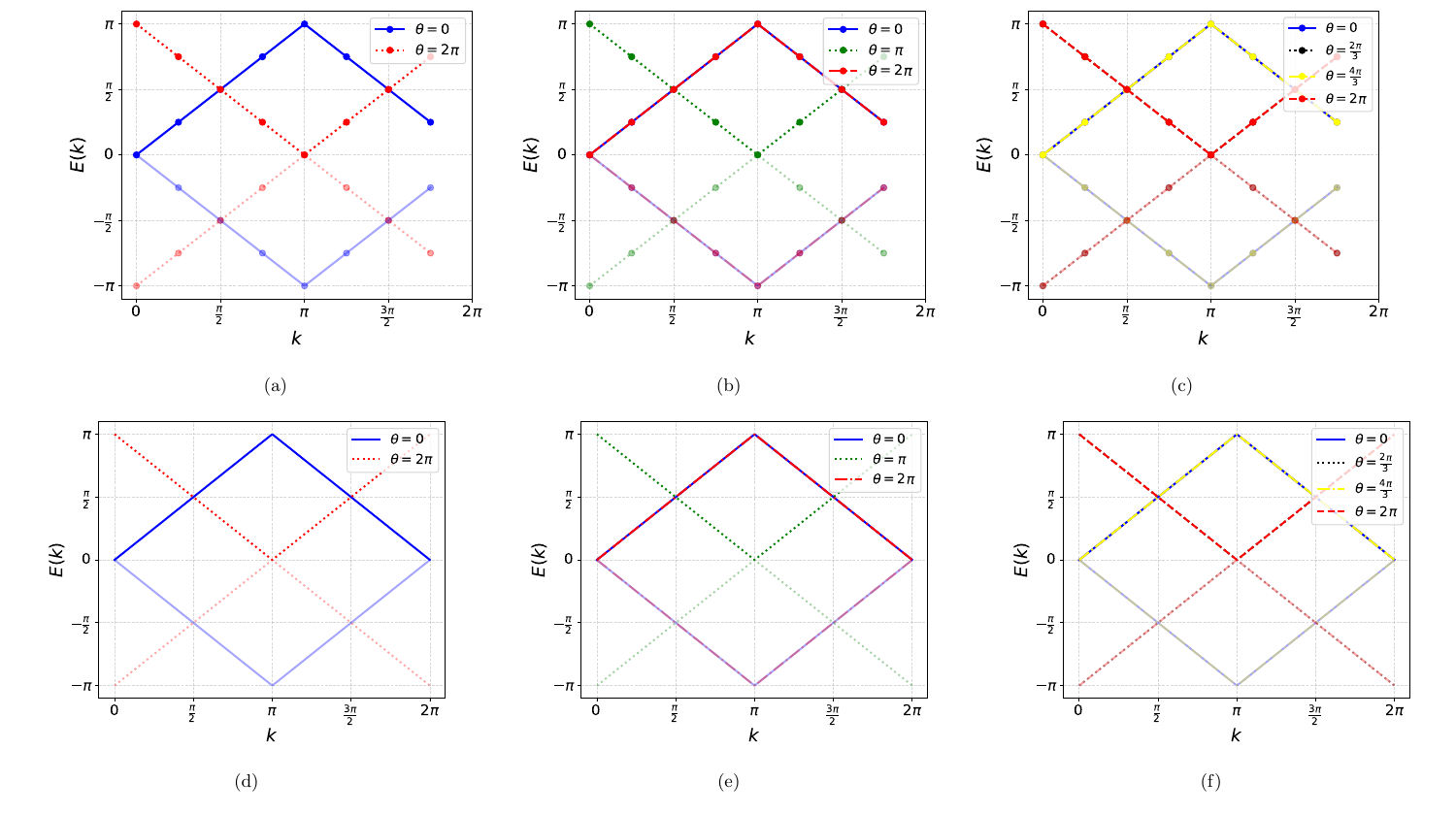}
\caption{\textbf{Energy gap closing (Dirac cones): } Energy dispersion relation vs quasi-momenta $k$ at rotation angle $\theta$ values (see Eq.~(17) of SM Sec. B) yielding gap closing on a $N=8$-cycle for (a) $T=1$  (b) $T=2$, (c) $T=3$; and that on a $N=1000$-cycle for (d) $T=1$ (e) $T=2$ and (f) $T=3$. One can control the energy gap closing and Dirac cone location by tuning the CQW parameters $T, \theta, N$.}
\label{2dplot}
\end{figure}

Furthermore, the 2D plots in Figs.~\ref{2dplot}(a)-(c) show the energy dispersion vs quasimomenta $k$, at the rotation angles $\theta$ derived in Eq.~\ref{eq31} with $T=1,2,3$, for both $N=8$-cycle and $N=100$-cycle. One can observe clearly the energy gap closing (Dirac cones). The number of gap closing points (Dirac cones) increases with increasing step-dependence parameter $T$. For $T=1$, two rotation angles $\theta=0, 2\pi$, for $T=2$, three rotation angles $\theta=0,\pi, 2\pi$ and for $T=3$ four rotation angles $\theta=0,\frac{2\pi}{3},\frac{4\pi}{3}, 2\pi$ (and so on), yield energy gap closing (Dirac cones). These results support the numerical results (3D plots) on energy gap closing at these $\theta$ values shown in Figs.~2 and 3 in the main and Figs.~\ref{f78t3}, \ref{ft1}-\ref{ft3}, and make them visually more traceable/pleasing.

\textbf{Theorem 1b: }
Gapped flat bands in CQW evolution arise at rotation angles which are odd multiples of $\frac{\pi}{T}$, where $T$ is the time-dependency parameter in CQW.

\textbf{\textit{Proof:}}
We find locations of the flat bands, where energy becomes independent of momentum $k$, i.e., group velocity $v_{gr}(k)=0$ (Eq.~(\ref{eq9})). Using energy dispersion in Eq.~(\ref{eq6}) or group velocity in Eq.~(\ref{eq9}), we get $v_{gr}(k)=0$, requires,
\begin{equation}
\begin{split}
   \cos \frac{T\theta}{2}=0
   \;\implies \theta=(2n+1)\frac{\pi}{T}, \;n\in \mathbb{Z_+} \cup \{0\}.
   \\&
\end{split}
\label{eq32}
\end{equation}
For instance, a step-independent CQW ($T=1)$ with $\theta=\pi$ and for a step-dependent CQW ($T=2$) with $\theta=\frac{\pi}{2}$, lead to the appearance of flat bands (gapped) with $E(k)=\pm \frac{\pi}{2}$, see Figs.~\ref{ft1}(c), \ref{ft2}(c), these are topological gapped flat-bands as the corresponding winding numbers are nonzero, i.e., $1$, see Fig.~\ref{ft1}(d-e-f) and Fig.~\ref{ft2}(d-e-f).

\begin{figure}[H]
\includegraphics[width = 18.5cm,height=4.5cm]{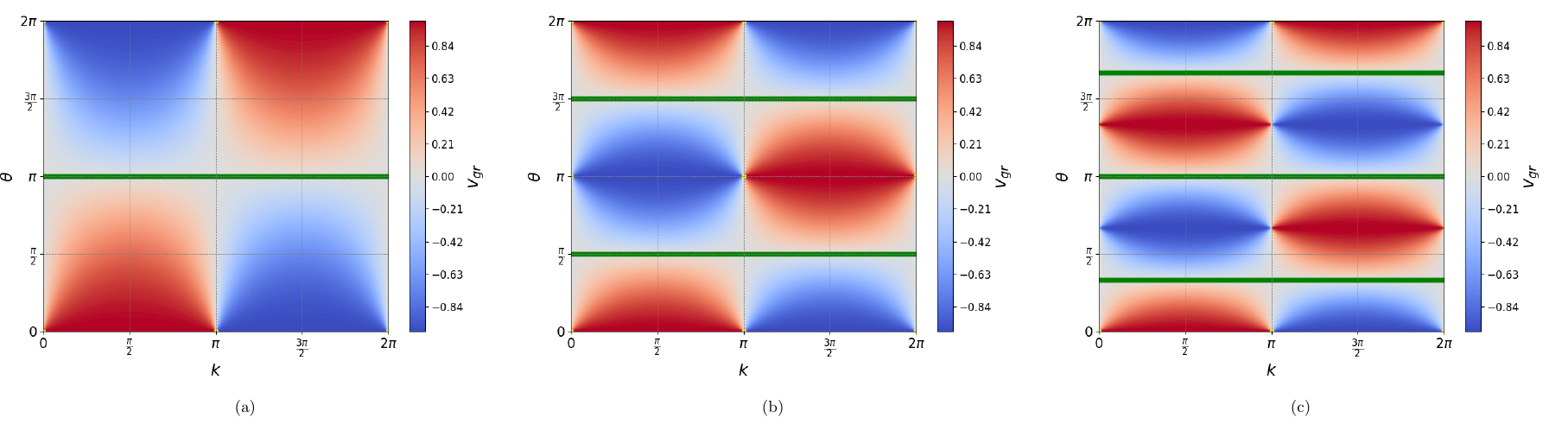}
\caption{Group velocity (Eq.~(\ref{eq9}) with + sign) vs quasi-momenta $k$ and rotation angle $\theta$ with (a) step-independent coin $T=1$, (b) step-dependent coin $T=2$, (c) step-dependent coin $T=3$, for CQW evolution. \textbf{Flat bands} (where energy $E(k)$ is independent of $k$) are signaled by zero group velocity ($v_{gr}=0, \forall k$), validating the condition in Eq.~(\ref{eq32}), see green dots.}
\label{grvel}
\end{figure}

The flat bands can also be verified from zero group velocity (Eq.~(\ref{eq9})) for all momentum ($k$) values, at specific rotation angles $\theta$, see Figs.~\ref{grvel}(a)-(c), validating the condition in Eq.~(\ref{eq32}). Further, since the effective mass (Eq.~(\ref{eq10})), $m^{*}(k,\theta,T)=\frac{1}{\frac{\partial v_{gr}}{\partial k}}$, the effective mass will become undefined (of the form $\frac{1}{0}$) whenever the group velocity ($v_{gr}$) is zero, i.e., at $\theta=(2n+1)\frac{\pi}{T}, \;n\in \mathbb{Z_+} \cup \{0\}$ (Eq.~(\ref{eq32})).
%at those points, undefined effective mass (Eq.~(\ref{eq10}))  with respect to momentum $k$, at specific $\theta$ values as derived in Eq.~(\ref{eq32}).Notably, undefined effective mass at rotation angles other than $\theta=(2n+1)\frac{\pi}{T}, \;n\in \mathbb{Z_+} \cup \{0\}$ (see Eq.~\ref{eq32}, e.g., $\theta=\pi$ for $T=1$ and  $\theta=\frac{\pi}{2},\frac{3\pi}{2}$ for $T=2$), are a result of nonzero constant group velocity or constant $\frac{ \partial E}{\partial k}$ and are not due to the occurrence of flat bands.

We note that gapless flat bands are not possible in CQW, as it would require $\cos \frac{T\theta}{2}\ne0$, which differs from the condition of flat band formation in CQW systems.

%and undefined effective mass (either of the form $\frac{0}{0}$ or $\frac{1}{0}$) for all $k$ values, at certain rotation angles $\theta$,
% \textbf{Note:} Undefined effective mass (yellow dots) at rotation angles other than $\theta=(2n+1)\frac{\pi}{T}, \;n\in \mathbb{Z_+} \cup \{0\}$ (flat band condition: Eq.~\ref{eq32}), are a result of nonzero constant group velocity or constant $\frac{ \partial E}{\partial k}$ or zero energy, and are not due to the occurrence of flat bands.

\textbf{Theorem 1c: }
    Rotational flat bands are only seen in even $4n$-cycles ($n\in\mathbf{N}$).
    
\textbf{\textit{Proof:}}
In the 4-cycle and 8-cycle cases see Figs.~\ref{ft1}(b) to \ref{f78t5}(b), the energy dispersion is found to be independent of $\theta$ at certain $k$-values, i.e., a flat band with respect to rotation angle $\theta$ (or, rotational flat band), which is not observed in 3,7-cycles, see Figs.~\ref{ft1}(a) to \ref{f78t5}(a). This observation is true for both step-dependent and step-independent coins, i.e., for any $T$ value.

Rotational flat bands are shown for the first time via this study, and they imply that the energy of the CQW system does not depend on the choice of the quantum coin/gate. In general, we can show that rotational flat bands manifest only for even cyclic graphs with the number of sites which are multiples of 4, i.e., $N=4,8,12,16,...$ From energy dispersion, in Eq.~(\ref{eq6}), both bands of energy $E(k)$ would be $\theta$ independent only if $\cos{k}=0$, or, $\cos{\frac{2\pi k'}{N}}=0$, i.e., when $k=(2n+1)\frac{\pi}{2}$, or,
\begin{equation}    
    k'=\frac{N}{2\pi}(2n+1)\frac{\pi}{2}=\frac{N}{4}(2n+1), 
\end{equation}
where $n=0,1,2,...$
Since $k'$ takes integer values only, the existence of rotational flat bands demands $N$ to be a multiple of 4. Thus, rotational flat bands only appear for $N=4,8,12,16$-cyclic graphs, i.e., even $4n$-cycles with $n\in\mathbf{N}$. Thus, rotational flat bands are absent for all odd-cycles like the 3, 5, 7-cycles and all even multiples of odd-numbered site-cyclic graphs like $N=6,10,14...-$cycles. See Figs.~\ref{f78t5}, \ref{ft1}, which clearly demonstrate that rotational flat bands manifest in 4, 8-cycles but not in 3,7-cycles, as predicted analytically.

%\textcolor{red}{We focus on generating and simulating topological phases and topological edge states which appear at the interface/boundary between two distinct topological phases (with different topological invariants: two winding numbers), e.g., see main text Fig. 1(b). Topological edge states are characterised by a large (or, near-unit) probability of occurrence, which is nearly independent of time, localised around the boundaries, as long as the initial state of the particle (quantum walker) has an overlap with the boundary states [20, 36, 44].}
 
%\textcolor{red}{We can analytically show where band closes with $E=0,\pm \pi$ in 3-cycle and 4-cycle as a fucntion of $\theta, N$--ONGOING}.

%\subsection{}

\onecolumngrid
\subsection{Generating topological edge states}
\begin{figure}[H]
\includegraphics[width = 18cm,height=4.2cm]{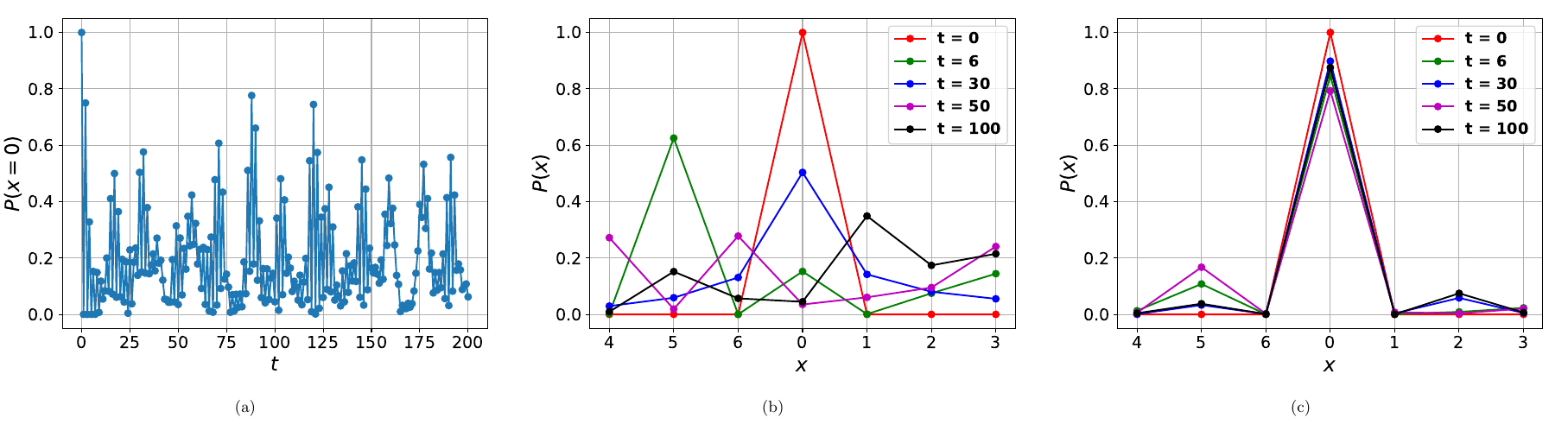}
\caption{(a) Probability of the particle at position $x=0$ vs time-step $t$ showing a non-periodic or chaotic CQW evolution (i.e., the particle does not return to its initial position through the time-evolution unlike in a periodic CQW evolution) with coin ($\theta=\frac{\pi}{3}$); (b) Absence of edge state due to identical topological phase ($\omega=1$) throughout position space i.e., no boundary; (c) Generation of edge state at the interface (site 0) between two distinct phases (i.e., with $\omega=-1$ and $\omega=+1$), via step-dependent CQW ($T=2$), for 7-cycle.}
\label{edge-7}
\end{figure}

One fascinating feature of topological phases is the ability to generate edge states, which appear at the interface between two distinct topological phases. Such topological edge states are characterised by large or near-unity probability at the boundary site, see  Fig.\ref{f0}(b), where the boundary is created by the site 0. To create edge states, we have numerous options of rotation angles and $T$ (see Figs.~\ref{ft2}-\ref{f78t5} and main text Fig.~2), for example, of the appearance of edge states, see main text Fig.~4 for 8-cycle and Fig.~\ref{edge-7} below for 7-cycle.

In Fig.~\ref{edge-7}, we consider step-dependent CQW ($T=2$, see main text Fig.~2) for a 7-cycle graph for which the position site 0 is acted on by coin ($\theta=\frac{7\pi}{5}$ ) with winding number $\omega= -1$, while the rest of the sites are acted on by coin ($\theta=\frac{\pi}{3}$) with winding number $\omega= +1$. This defines a boundary at site 0, see main text Fig.~1(b). As in the main text, we consider the initial state of the quantum walker, $\ket{\psi(0)}=\ket{0}_p\otimes\frac{\ket{0}_c+\ket{1}_c}{\sqrt{2}}$. Significant values of probability at site 0 due to the overlap of the walker's initial site with the boundary are characteristic of an edge state~\cite{kita-pra,pxue,chandra-topo}. Methods using split-step and split-coin operators (resource-consuming) to create edge states on 1D line have been shown in Refs.~\cite{kita-pra,pxue,chandra-topo}. Herein, with a step-dependent coin in CQW, we observe long-lived edge states (persistent over time $t$) for 7-cycle in Fig.~\ref{edge-7} and also for 8-cycle (see main text Fig.~4). For the first time, we obviate the need to use split-step or split-coin quantum walks to create edge states, and we use only experimentally resource-saving small cyclic graphs.
We can generate numerous/infinite such topological edge states by creating a boundary between two distinct phases in all odd or even cyclic graphs and for $T>2$ values, too.

\onecolumngrid
\begin{figure}[H]
\includegraphics[width = 17cm,height=5cm]{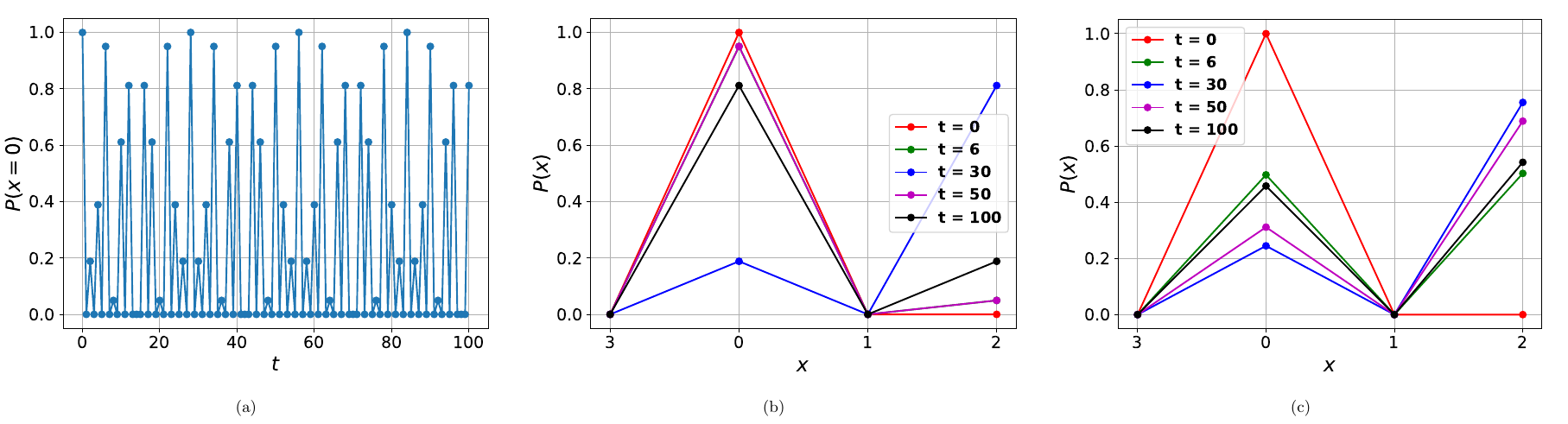}
\caption{(a) Probability of the particle at position $x=0$ vs time-step $t$ showing periodic evolution with $\theta=\frac{\pi}{7}$; (b) Absence edge state due to same topological phase with $\theta=\frac{\pi}{7}$, throughout the position space i.e., no phase boundary; (c) Attempt to generate of edge state generation at the interface ($x=0$) between two distinct phases (i.e., with $\omega=-1,\theta=\frac{3\pi}{2}$for site 0 and $\omega=+1, \theta=\frac{\pi}{7}$ for all remaining sites), via periodic CQW, for 4-cycle. No clear sign of topological edge state in (c) at the boundary site, possibly masked by periodicity.}
\label{edge-4-periodic}
\end{figure}

\subsubsection{CQW periodicity and very-small cycle graphs can mask edge state formation} 
\begin{figure}[H]
\includegraphics[width = 17cm,height=5cm]{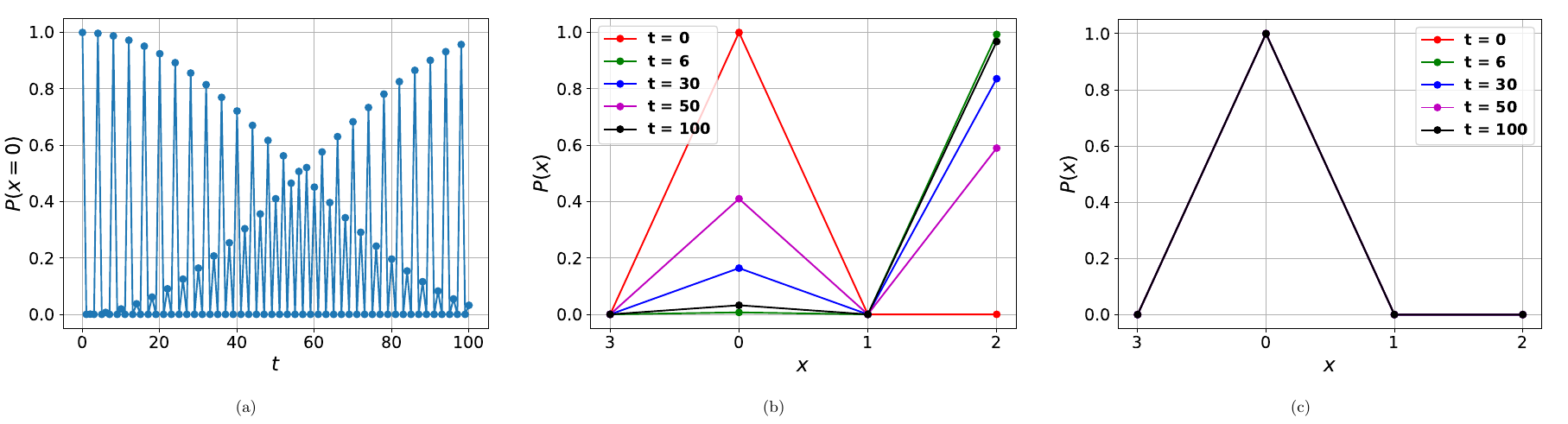}
\caption{(a) Probability of the particle at position $x=0$ vs time-step $t$ showing non-periodic (chaotic) evolution with $\theta=\frac{\pi}{113}$; (b) Absence edge state due to same topological phase with $\theta=\frac{\pi}{113}$ throughout the position space i.e., no phase boundary; (c) Generation of edge state generation at the interface ($x=0$) between two distinct phases (i.e., with $\omega=1,\theta=\frac{\pi}{113}$ for site 0 and $\omega=-1, \theta=\frac{3\pi}{2}$ for all remaining sites), via CQW, for 4-cycle.}
\label{edge-4}
\end{figure}

We observe edge states clearly in chaotic (non-periodic) CQWs and 7,8-cycles ($N=7,8$), see Figs.~\ref{edge-7} and Fig.~4 in main. CQW periodicity and very-small cycle graphs (e.g., 3,4-cycles) may mask edge states at the boundary sites due to periodic evolution of the walker's initial site and repeated superposition/interference, see Figs.~\ref{edge-4-periodic}-\ref{edge-4}.
In Fig.~\ref{edge-4-periodic}(c), we consider step-dependent CQW ($T=2$, see Fig.~2 in main) for a 4-cycle graph for which the position site 0 is acted on by coin ($\theta=\frac{\pi}{7}$ ) with winding number $\omega= 1$, while the rest of the sites are acted on by coin ($\theta=\frac{3\pi}{2}$) with winding number $\omega= -1$. This defines a boundary at site 0. As in the main, we consider the initial state of the quantum walker, $\ket{\psi(0)}=\ket{0}\otimes\frac{\ket{0}_c+\ket{1}_c}{\sqrt{2}}$. The two coins with $\theta=\frac{\pi}{7}$ and $\theta=\frac{3\pi}{2}$ individually yield periodic evolution and we do not get any clear sign of topological edge state in this case. Similarly, in Fig.~\ref{edge-4}(c), we choose $\theta=\frac{\pi}{113}$ for site 0 and $\theta=\frac{3\pi}{2}$ for other sites, where individually $\theta=\frac{\pi}{113}$ yields a chaotic CQW (upto $t=100$) while $\theta=\frac{3\pi}{2}$ yields periodic CQW, and we observe an edge state in this case. Thus, we use non-periodic (chaotic) CQW as well as cycles larger than 4-cycle, like 7,8-cycles to observe edge states clearly, see Fig.~\ref{edge-7} and Fig.~4 in main. Below, we prove the robustness of topological edge states with an example 8-cycle against static and dynamic disorder and phase-preserving perturbations in the following section.

%Then in Sec.~C, we show additional results on edge states on odd and even cycles. Moreover, we prove the robustness of of topological edge states against small static and dynamic disorder in quantum gate (coin) operations, in Sec.~D. Then Sec.~E shows the resilience of edge states against phase-preserving perturbations.
\subsection{Effect of disorder on topological edge states}

\subsubsection{Under static coin disorder}
\begin{figure}[H]
\includegraphics[width = 18cm,height=4.2cm]{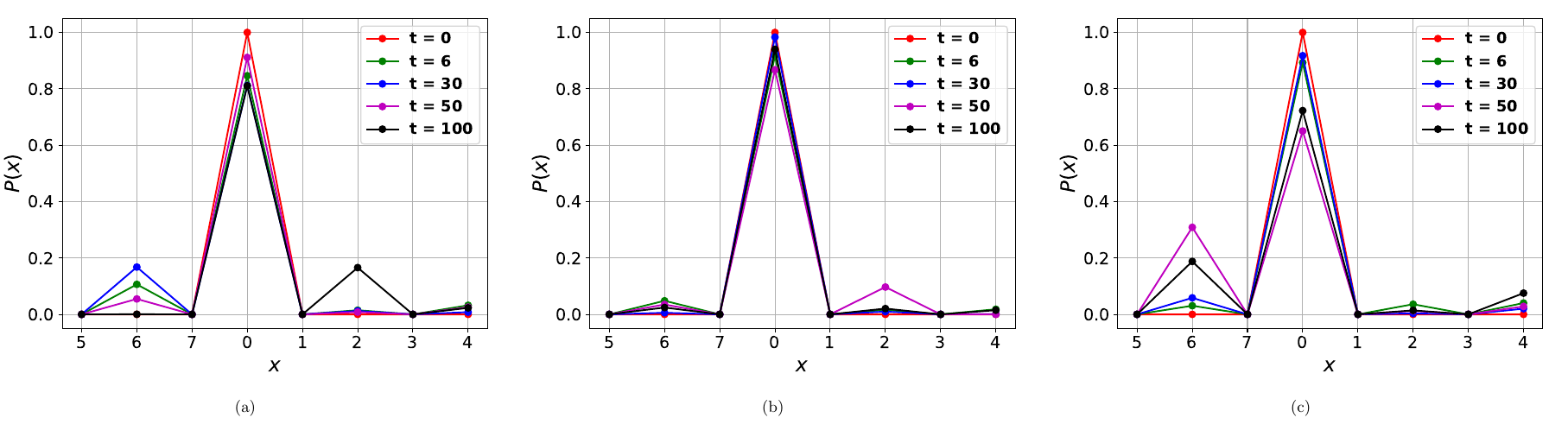}
\caption{(a) Edge state at the interface (site 0) between two distinct phases (i.e., with $\theta=\frac{7\pi}{5},\;\omega=-1$ and $\theta=\frac{\pi}{3},\;\omega=+1$), via step-dependent CQW ($T=2$) on a 8-cycle lattice, without static disorder, i.e., $\Delta_d=0$. (b) Effect of static coin disorder of strength $\Delta_s=0.1$ on the edge state shown in (a). (c) Effect of static coin disorder of strength $\Delta_s=0.2$ on the edge state shown in (a). In (b)-(c), 500 disorder realizations are taken and the probability $P(x)$ is averaged over the 500 realizations.}
\label{edge-8-stat}
\end{figure}

We now consider static coin disorder in the CQW evolution with disorder strength $\Delta_s$. This implies every site $(x)$ dependent rotation angle ($\theta(x)$) used for generating topological phases and edge states changes as, $\theta(x)\rightarrow \theta(x)+\Delta_s\delta\theta(x)$ and the random numbers $\delta\theta(x) \in [-\pi,\pi]$ with size same as the number of sites on the cyclic graph, are drawn from an uniform distribution~\cite{p5,sasha} for every random realization. Clearly, $\Delta_s=0$ refers to no static disorder in CQW. Fig.~4 in main and in Fig.~\ref{edge-8-stat}(a), we discussed the generation edge state at site 0 via creating a phase boundary with two distinct phases with rotation angles $\theta=\frac{7\pi}{5}$ (assigned to site 0, with winding number -1) and $\theta=\frac{\pi}{3}$ (assigned to other sites, with winding number +1), see also Fig.~\ref{edge-8-stat}(a), where we see significant probability amplitude at the site 0 which is persistent over time $t$ too (long-lived state). 
In Fig.~\ref{edge-8-stat}(b) and Fig.~\ref{edge-8-stat}(c), we show the effect of static coin disorder of strengths: $\Delta_s=0.1$ and $\Delta_s=0.2$, respectively, on the edge state shown in Fig.~\ref{edge-8-stat}(a). We observe that the edge state is robust against small static disorder of strength $0<\Delta_s\lesssim 0.2$, as the probability of the particle at site 0 does not decay significantly under the small disorder strengths. However, on increasing the disorder strength $\Delta_s\gtrsim 0.2$, the edge state amplitude gets affected by the disorder significantly.

\onecolumngrid
\subsubsection{Under dynamic coin disorder}
We consider dynamic coin disorder in the CQW evolution with disorder strength $\Delta_d$. This implies every site $(x)$ dependent rotation angle ($\theta(x)$) used for generating topological phases and edge states changes as, $\theta(x)\rightarrow \theta(x)+\Delta_d\delta\theta(t)$ and the site-independent random numbers $\delta\theta(t) \in [-\pi,\pi]$ with size same as the number of time-steps, are drawn from an uniform distribution~\cite{p5}, for a specific random realization. We consider 500 such random realizations in order to estimate the quantity of interest, i.e., the probability $P(x)$ of finding the walker at site $x$, also see Ref.~\cite{p5} for more details on disorder realization in CQW evolution.

\begin{figure}[H]
\includegraphics[width = 18cm,height=4.2cm]{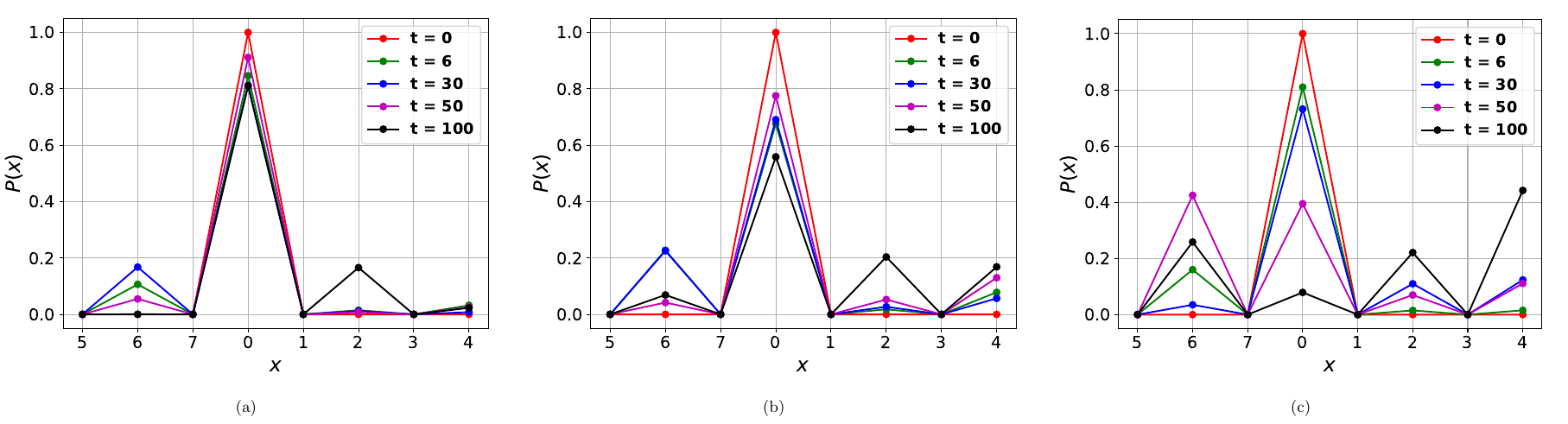}
\caption{(a) Edge state at the interface (site 0) between two distinct phases (i.e., with $\theta=\frac{7\pi}{5},\;\omega=-1$ and $\theta=\frac{\pi}{3},\;\omega=+1$), via step-dependent CQW ($T=2$) on a 8-cycle lattice, without dynamic disorder, i.e., $\Delta_d=0$. (b) Effect of dynamic coin disorder of strength $\Delta_d=0.05$ on the edge state shown in (a). (c) Effect of dynamic coin disorder of strength $\Delta_d=0.1$ on the edge state shown in (a).  In (b)-(c), 500 disorder realizations are taken and the probability $P(x)$ is averaged over the 500 realizations.}
\label{edge-8-dyn}
\end{figure}

$\Delta_d=0$ refers to no dynamic disorder in the system and CQW evolution. Fig.~4 in main discusses the generation edge state at site 0 via creating a phase boundary with two distinct phases with rotation angles $\theta=\frac{7\pi}{5}$ (at site 0, with winding number -1) and $\theta=\frac{\pi}{3}$ (at other sites, with winding number +1), see Fig.~\ref{edge-8-dyn}(a), where we see significant probability amplitude at site 0 which is persistent over time $t$ too (long-lived). 
In Fig.~\ref{edge-8-dyn}(b), Fig.~\ref{edge-8-dyn}(c), we see the effect of dynamic coin disorder of strengths, $\Delta_d=0.05$ and $\Delta_d=0.1$, respectively, on the topological edge state shown in Fig.~\ref{edge-8-dyn}(a). We observe that the edge state is robust against very small dynamic disorders of strength $0<\Delta_d\le 0.05$, as the probability of the particle at site 0 does not decay significantly under the small dynamic disorder strengths. However, on increasing the disorder strength $\Delta_d >0.05$, the edge state amplitude gets affected by the disorder significantly.
    
\onecolumngrid
\subsubsection{Under phase-preserving perturbations}
\begin{figure}[H]
\includegraphics[width = 18.6cm,height=4.5cm]{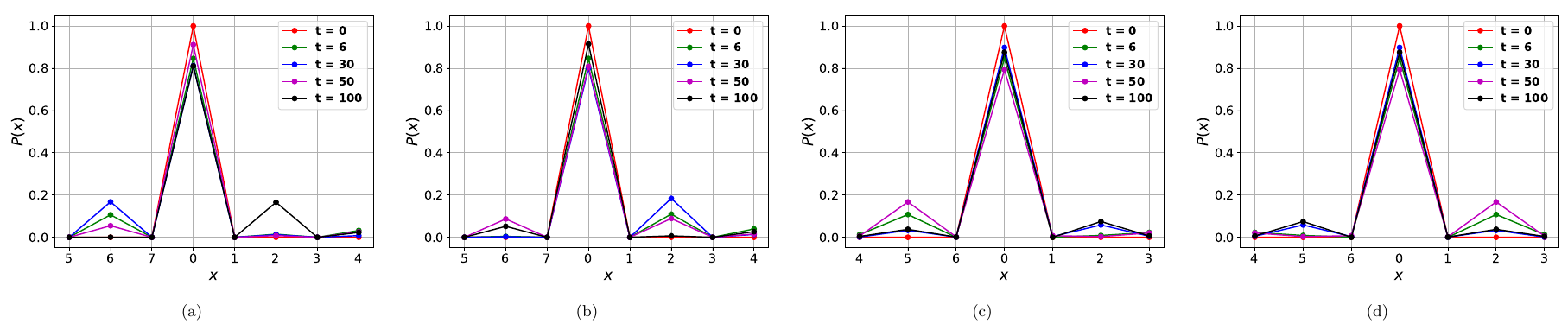}
\caption{(a) Absence of edge state due to identical topological phase ($\theta=\frac{\pi}{3}\;,\omega=1$) throughout position space i.e., no boundary; (a) Generation of edge state at the interface (site 0) between two distinct phases (i.e., with $\theta=\frac{7\pi}{5},\;\omega=-1$ and $\theta=\frac{\pi}{3},\;\omega=+1$), via step-dependent CQW ($T=2$); (b) Persistence of edge state at the interface (site 0) between two distinct phases (i.e., with $\theta=\frac{8\pi}{5},\;\omega=-1$ and $\theta=\frac{\pi}{3},\;\omega=+1$), via step-dependent CQW ($T=2$); for 8-cycle. (c) Generation of edge state at the interface (site 0) between two distinct phases (i.e., with $\theta=\frac{7\pi}{5},\;\omega=-1$ and $\theta=\frac{\pi}{3},\;\omega=+1$), via step-dependent CQW ($T=2$); (d) Persistence of edge state at the interface (site 0) between two distinct phases (i.e., with $\theta=\frac{8\pi}{5},\;\omega=-1$ and $\theta=\frac{\pi}{3},\;\omega=+1$), via step-dependent CQW ($T=2$); for 7-cycle.}
\label{edge-7-8-pert-1}
\end{figure}

Phase-preserving perturbations~\cite{pxue} are introduced by modifying the rotation angles ($\theta$) used for generating topological phases and edge states in the CQW evolution, without changing the topological number. This is accomplished as follows, say $\theta=\frac{7\pi}{5}$  and $\theta=\frac{8\pi}{5}$ have the same winding number $\omega=-1$ and then we permute $\theta$ from $\frac{7\pi}{5}$ to $\frac{8\pi}{5}$ without changing the associated winding number (topological invariant). Fig.~4 in main and Fig.~\ref{edge-7-8-pert-1}(a) for a 8-cycle, we discussed the generation edge state at position site 0 via creating a phase boundary with two distinct phases with rotation angles $\theta=\frac{7\pi}{5}$ (assigned to site 0, winding number -1) and $\theta=\frac{\pi}{3}$ (assigned to other sites, winding number +1), where we see significant probability amplitude at the site 0 which is persistent over time $t$ too (long-lived).
In Fig.~\ref{edge-7-8-pert-1}(b), we introduce the perturbation via the rotation angle assigned to site 0 is modified to  $\theta=\frac{8\pi}{5}$, but with the same winding number -1. We observe that the edge state is robust against these topological phase-preserving perturbations, as the probability of the particle at site 0 does not decay due to the perturbation. 
%However, on increasing the disorder strength $\Delta_d\gtrsim 0.1$ , the edge state amplitude get affected by the disorder significantly.
%\begin{widetext}

The same robustness of edge state is observed for 7-cycle too, see Fig.~\ref{edge-7-8-pert-1}(c-d) and can be shown for cyclic graphs with arbitrary $N$-sites.

%\end{widetext}

\onecolumngrid

\subsection{Change in initial states and its effect on topological edge states via SD-CQW}

Herein, we demonstrate that the step-dependent CQW (SD-CQW) scheme for generating edge states is state-independent, i.e., it is completely independent of the nature (superposed, out-of-phase superposed, uneven superposed and unsuperposed) of the initial coin state (initial state) of the quantum walker.

\begin{figure}[H]
\includegraphics[width = 18.6cm,height=4.5cm]{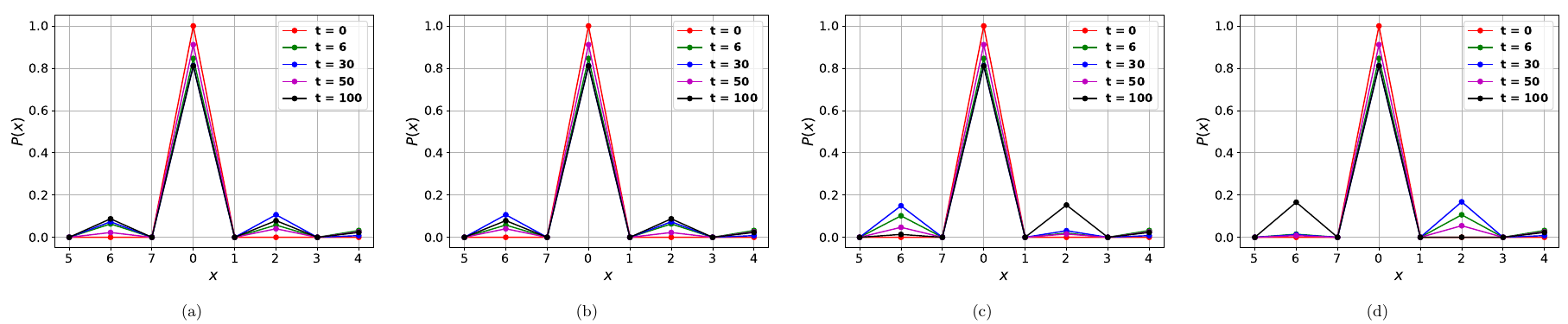}
\caption{\textbf{State-independence:} Generation of edge state at the interface (site 0) between two distinct phases (i.e., with $\theta=\frac{5\pi}{2}, \omega=-1$ and $\theta=\frac{\pi}{3}, \omega=+1$), via step-dependent CQW ($T=2$) on 8-cycle for initial state $\ket{\psi(0)}$: (a) $\ket{\psi(0)}=\ket{0}_p\otimes\ket{0}_c$ with non-superposed coin state, (b)  $\ket{\psi(0)}=\ket{0}_p\otimes\ket{1}_c$, with non-superposed coin state, (c) $\ket{\psi(0)}=\ket{0}_p\otimes\frac{\sqrt{3}\ket{0}_c+\ket{1}_c}{2}$ with uneven superposition in the coin state, and (d) $\ket{\psi(0)}=\ket{0}_p\otimes\frac{\ket{0}_c-\ket{1}_c}{\sqrt{2}}$ with out-of-phase superposed coin state. Here, subscript $p$ refers to the position state and subscript $c$ refers to the coin state of the quantum walker.}
\label{statedep}
\end{figure}

Our approach to generating topological phases and edge states is state-independent, i.e., it applies to any kind of initial states of the quantum walker, $\ket{\psi(0)}$, ranging from non-superposed coin states, unequal superposition in the coin state, equally superposed coin state and out-of-phase superposed coin state. The influence of different arbitrary initial states on the edge states within our SD-CQW scheme is illustrated in Figs.~\ref{statedep}(a)–\ref{statedep}(d). In Figs.~\ref{statedep}(a)–\ref{statedep}(d), we consider four different types of initial states  $\ket{\psi(0)}$ of the quantum walker, such as non-superposed coin states: $\ket{\psi(0)}=\ket{0}_p\otimes\ket{0}_c,\; \ket{\psi(0)}=\ket{0}_p\otimes\ket{1}_c$, unequal superposition in the coin state:  $\ket{\psi(0)}=\ket{0}_p\otimes\frac{\sqrt{3}\ket{0}_c+\ket{1}_c}{2}$ and out-of-phase superposed coin state: $\ket{\psi(0)}=\ket{0}_p\otimes\frac{\ket{0}_c-\ket{1}_c}{\sqrt{2}}$. 

%\begin{widetext}

The remaining type of initial state with the equal superposition in the coin state, i.e.,  $\ket{\psi(0)}=\ket{0}_p\otimes\frac{\ket{0}_c+\ket{1}_c}{\sqrt{2}}$, is already considered in Main Fig.~3(c) and it shows edge state, which is indicated by the significant values of probability at position site 0. It is evident from Figs.~\ref{statedep}(a)–\ref{statedep}(d), that all the other three types of initial states also yield edge states over all time steps identical to Main Fig.~3(c). This shows that the generation of the edge state via our SD-CQW scheme is completely independent of the nature of the initial coin state or, the initial state of the quantum walker particle. This proves the state independence of our scheme.
%\textcolor{red}{In Sec.~\textcolor{blue}{E}, we demonstrate that our step-dependent CQW (SD-CQW) scheme for generating edge states is state-independent i.e., it is completely independent of the nature (superposed, out-of-phase superposed, uneven superposed and unsuperposed) of the initial coin state or the initial state of the quantum walker particle. We compute the resource overhead (operator count scaling) of our SD-CQW scheme, in Sec.~\textcolor{blue}{F}, and therein we show that our scheme has an exponential advantage in its experimental implementation as compared to the existing protocols.}

\subsection{Resource overhead of our SD-CQW scheme vs. existing schemes}

The shift operators and coin (gate) operators in a quantum walk evolution are respectively the polarizing-beam-splitters (PBS or Jones plates) and waveplates in a photon-based implementation~\cite{bian,pxue,karimi19}. After the quantum evolution, the resulting probability distribution is read out using single‑photon detectors (which are expensive too). Analogous devices will mimic the implementation of our scheme in other platforms, like trapped ions, cold atoms, NMR systems~\cite{perets, peruzzo, schreiber, broome, tamura, schmitz, zaehringer, ryan}. The number of coin and shift operators, along with the number of particle detectors, constitutes the resource overhead (operator count plus detector count) of a quantum walk protocol. Herein, we quantify the resource overhead of our SD-CQW scheme (i.e., quantum walk on cyclic graphs or small periodic lattices) and existing split-step (SS-QW) and split-coin (SC-QW) quantum walk protocols~\cite{kita-pra,pxue,chandra-topo,barry-pra} in nonperiodic 1D lattices/lines, in order to generate topological phases and edge states, in an experiment (e.g., photonic implementation). Our scheme demonstrates a significant reduction in experimental resource overhead, requiring fewer coin and shift operators as well as minimal (fixed) single-photon detectors compared to SS-QW and SC-QW protocols. Quantitatively, our SD-CQW scheme achieves a twofold decrease in operator count and employs a constant number of particle detectors (as opposed to the O$(2\tau)$ scaling in existing protocols for $\tau$ time steps), thereby enhancing the feasibility of large-scale implementations of our CQW approach for topological phase and edge state generation in photonic and other quantum platforms. See Table~\ref{reso} for a clear juxtaposition of the total resource overhead across different quantum walk protocols. Below, we prove this significant achievement of our scheme.
\begin{table}[h!]
    \centering
    \scriptsize
    \resizebox{\columnwidth}{!}{%
    \renewcommand{\arraystretch}{1.5}
    \begin{tabular}{|c|c|c|c|c|c|}
       \hline        
       \textbf{Protocol} & \textbf{\makecell{No. of distinct coins \\required to construct \\phase boundary\\($\#_{\theta}$)} }& \textbf{\makecell{No. of\\ detectors\\($\#_{d}$)}}& \textbf{\makecell{No. of shift \\operations\\($\#_{s}$)}} &\textbf{\makecell{No. of coin \\operations\\($\#_{c}$)}}& \textbf{\makecell{Resource overhead \\($R=\#_{d}+\#_{s}+\#_{c}+\#_{\theta}$)}}\\
        \hline
        \multicolumn{6}{|c|}{\textbf{CQW: Quantum walks on cyclic graphs (for $N$, e.g., 7, 8,...)}} \\ \hline
        \rowcolor{gray!15} SD-CQW (this work) & $2$ & $N$ & $\tau$& $\tau$& $R_{SD}=2\tau+N+2$ \\ \hline
         \multicolumn{6}{|c|}{\textbf{Example: CQW: Quantum walks on cyclic graphs ($N=8,\; \tau=100$)}} \\ \hline
        \rowcolor{gray!15}SD-CQW (this work) & $2$ & $8$ & $100$&$100$& $R_{SD}=210$ \\ \hline
        \multicolumn{6}{|c|}{\textbf{QW: Quantum walks on 1D lattices (for $N_L$, e.g., 100, 200, 1000,...)}} \\ \hline
        SS-QW & $3$ & $(2\tau+1)$ & $2\tau$& $2\tau$& $R_{SS}=4\tau+(2\tau+1)+3$\\ \hline
        SC-QW & $3$ & $(2\tau+1)$ & $2\tau$& $2\tau$& $R_{SC}=4\tau+(2\tau+1)+3$  \\ \hline
         \multicolumn{6}{|c|}{\textbf{Example: QW: Quantum walks on 1D lattices ($N_L=100,\;\tau=100$)}} \\ \hline
        SS-QW & $3$ & $201$ & $200$&$200$& $R_{SS}=604$ \\ \hline
        SC-QW & $3$ & $201$ & $200$&$200$& $R_{SC}=604$   \\ \hline 
    \end{tabular}
    }
    \caption{Comparison of resource overhead in implementing different quantum walk protocols, i.e., ours: CQW on cyclic graphs (SD-CQW) vs. existing: QW on nonperiodic infinite 1D lattices (SS-QW and SC-QW), to generate topological phases and edge states. {\small{Note: $N_L$ denotes the number sites on 1D lattice and $N$ is that on a cyclic lattice/graph, while $\tau$ denotes the time steps in quantum walk dynamics. This estimate is done keeping a photon-based experimental implementation in mind; this will change accordingly if we implement on a quantum computer, but the excellent resource-saving nature of SD-CQW (our protocol) will be evident there, too, and in fact, the efficiency of SD-CQW will be much higher for quantum computer implementation.}}}
    \label{reso}
\end{table}

The time evolution of a quantum particle (quantum walker) via our step-dependent CQW (SD-CQW) scheme is governed by the evolution operator, $U = \hat{S} \hat{C_2(\theta, T)}$. Whereas, the time-evolution of quantum particles via split-step quantum walks (SS-QW) and split-coin quantum walks (SC-QW) on nonperiodic 1D lattices, are governed respectively by the evolution operators,
\begin{equation}
\text{for SS-QW: } U_{SS} = \hat{S}_{+} \hat{C}(a) \hat{S}_{-} \hat{C}(b) \text{ and for SC-QW: } U_{SC} = \hat{S_f} \hat{C}(a) \hat{S_f} \hat{C}(b),
\label{evo1}
\end{equation}
where the shift operators are: 

$\hat{S} = \sum_{q=0}^{1}\sum_{x=0}^{N-1}\ket{(x+(-1)^{q}) \text{ mod } N}\bra{x}\otimes\ket{q_c}\bra{q_c},$ for SD-CQW; while

\[
\hat{S}_{+} = \sum_{n=-\infty}^{\infty} \ket{0}_c \bra{0}_c \otimes \left( \ket{x+1} \bra{x} + \ket{1} \bra{1} \otimes \ket{x} \bra{x} \right),\;\;
\hat{S}_{-} =  \sum_{n=-\infty}^{\infty} \ket{0}_c \bra{0}_c \otimes \left( \ket{x} \bra{x} + \ket{1} \bra{1} \otimes \ket{x-1} \bra{x} \right),
\] for SS-QW, and 
 $\hat{S}_{f} = \sum_{n=-\infty}^{\infty} \ket{0}_c \bra{0}_c \otimes \left( \ket{x+1} \bra{x} + \ket{1} \bra{1} \otimes \ket{x-1} \bra{x} \right),$ for SC-QW. \\The coin operators have the generic form, $\hat{C}(\theta) \equiv e^{-i \frac{\theta}{2} \sigma_y},$ and $\hat{C}_2(\theta, T) \equiv e^{-i \frac{T\theta}{2} \sigma_y}.$
If $\ket{\psi(0)}$ represents the initial state of the quantum walker, then the evolved quantum state at time-step $t$ is, $\ket{\psi(t)} = U^t \ket{\psi(0)}$.

At time $t=\tau$, the evolution operators of SD-CQW ($U$ in our protocol), SS-QW ($U_{SS}$), and SC-QW ($U_{SC}$), take the forms: 
\begin{equation}U^\tau=(\hat{S}\,\hat{C}_2(\theta, T))^\tau,\;\;U_{SS}^\tau = (\hat{S}_{+}\,\hat{C}(a)\,\hat{S}_{-}\,\hat{C}(b))^\tau,\;\;\text{and }U_{SC}^\tau = (\hat{S_f} \hat{C}(a) \hat{S_f} \hat{C}(b))^\tau.\label{opr}\end{equation}
%One can compute theor gate count scaling (number of elementary gates or unitary blocks required) of $2\tau$ in our SD-CQW scheme, compared to $4\tau$ in both SS-QW and SC-QW protocols, to establish the quantum walk dynamics and to generate edge states in real space by creating phase boundaries. Our scheme is therefore less resource-consuming, achieving an exponential reduction in gate count scaling relative to SS-QW and SC-QW protocols.%%

One can compute the resource overhead i.e., the operator count scaling (number of operators required) in closed form and the detector count (number of particle detectors). Here operators include coin operators and shift operators, and in each time step of our protocol (SD-CQW), there are 2 operators, and in each time step of SS-QW (or SC-QW), there are 4 operators, see Eq.~(\ref{opr}) above. Thus, for a quantum walk protocol after $\tau$ time steps, the operator count scales as, \begin{equation}
G_{\rm SD}(\tau) = 2\tau ~\text{(for SD-CQW, our protocol)},~~
G_{\rm SS/SC}(\tau) = 4\tau~\text{(for SS-QW or SC-QW protocols)}.
\label{gs}
\end{equation}
The split-step and split-coin protocols demand a multiplicative overhead as compared to our SD-CQW protocol, i.e.,
\begin{equation}\frac{G_{\rm SS/SC}(\tau)}{G_{\rm SD}(\tau)}=\frac{4\tau}{2\tau}=2,\end{equation}
or, there is $2$ times greater operator count (resource) requirement than our scheme. The absolute operator resource saving count of our scheme (SD-CQW) is,
\begin{equation}
  \Delta G(\tau)=4\tau-2\tau=2\tau,  
\end{equation}
while the relative saving in terms of operator count is,
\begin{equation}  
\frac{\Delta G(\tau)}{4\tau}=50\%. 
\end{equation}

In addition to this significant reduction in operator count by a factor of two, our SD-CQW protocol also achieves substantial savings in the particle-detector requirement. Specifically, our SD-CQW scheme requires a constant number of detectors equal to the number of sites ($N=7,8,\dots$) on the small cyclic (periodic) graph, irrespective of the number of time steps in the quantum walk evolution. In contrast, the SS-QW and SC-QW protocols demand $(2\tau+1)$ detectors for $\tau$ time steps, resulting in a linearly growing resource overhead in the detector count. Thus, our scheme exhibits an $O(1)$ scaling in detector count, as opposed to the $O(2\tau)$ scaling in existing protocols, thereby greatly enhancing the scalability and experimental feasibility of topological phase and edge-state generation using SD-CQW on photonic and other quantum platforms. In addition, our SD-CQW scheme employs two distinct coins (i.e., $\hat{C_2(\theta, T)}$ with two different $\theta$ values) to construct a phase boundary (necessary for edge state generation) on the real position space. A coin with step dependency in our scheme, such as $\hat{C}_2(\theta, T=2)$ (which appears to be a product of two coins $\hat{C_2}(\theta)\cdot\hat{C_2}(\theta)$) is equivalently a single coin, i.e., a single waveplate with a fixed rotation degree, in a photonic setup. In contrast, SS-QW or SC-QW requires three distinct coins (i.e., $\hat{C(b)}$ with two different $b$ values and $\hat{C(a)}$ with a fixed $a$ value) to construct a similar phase boundary in real position space, in order to generate an edge state. The overall resource overhead of our SD-CQW scheme is \boxed{R_{\rm SD}=G_{\rm SD}+N=2\tau+N+2}, compared to the significantly larger \boxed{R_{\rm SS/SC}=G_{\rm SS/SC}+(2\tau+1)+3=4\tau+(2\tau+1)+3} in the SS-QW and SC-QW protocols, as summarized in Table~\ref{reso}.

\underline{As a concrete example:} for a cyclic quantum walk with $N=8$ sites and $\tau=100$ time steps, our SD-CQW scheme requires a total resource overhead of $R_{\rm SD}=2\tau+N+2=210$, whereas both SS-QW and SC-QW protocols require $R_{\rm SS/SC}=4\tau+(2\tau+1)+3=604$. This demonstrates that our approach reduces the total experimental resource requirement by more than a factor of \textbf{$2.88$}, corresponding to an absolute saving of $394$ devices (detectors plus operators) or a relative saving of approximately \textbf{$65.23\%$}, while maintaining identical dynamical capability in generating topological phase boundaries and edge states. Such a drastic reduction in operational and detection resources underscores the superior experimental efficiency and practical implementability of our unique SD-CQW protocol.

%As a concrete example, at time step $\tau=8$, we have $G_{\rm SD}=2^8=256$ operators in our protocol whereas, $G_{\rm SS/SC}=4^8=65\,536$ operators in SS-QW or SC-QW protocol, assuming the same number of lattice sites. Thus, the latter uses a factor of $2^8=256$ times more operators, i.e., an absolute saving of $65\,280$ operators and a relative saving of $1-2^{-8}\approx99.61\%$ can be achieved via our SD-CQW scheme. For $\tau=10$ the multiplicative operator count (resource) overhead grows to $2^{10}=1024$ (with $G_{\rm SD}=1024$ vs.\ $G_{\rm SS/SC}=1\,048\,576$), illustrating the exponential advantage of our SD-CQW scheme.

%he multiplicative overhead further increases to $2^{10}=1024$ (with $G_{\rm SD}=1024$ vs.\ $G_{\rm SS/SC}=1{,}048{,}576$), clearly demonstrating the exponential advantage and resource efficiency of our SD-CQW framework over the SS-QW and SC-QW schemes.

\subsection{Algorithms and Python codes}
Below, we put forth two algorithms: (1) to generate edge states in cyclic graphs via CQW dynamics and (2) to realize the effects of disorder of the edge states. We also provide typical Python codes to visualize them in GitHub~\cite{git} (email us for permission to access the codes).
\subsubsection{Algorithm for generating edge states via step-dependent CQW on cyclic graphs}

\begin{algorithmic}[1]
\Require \# of time steps $J$, \# of sites $N = 8$ (say for a 8-cycle graph), time-dependency parameter $T$
\Require Coin parameters i.e., topological rotation angles $\theta_i$ which give non-zero winding numbers
\Ensure Probability distribution $\texttt{pro}[t][x]$ for $t=0$ to $J$, index to denote sites: $x=0$ to $N-1$

\State Initialize probability array: $\texttt{pro} \leftarrow$ zeros of shape $(J+1, N)$
\State Initialize the walker's state $\ket{\psi(0)}=\ket{0}\otimes\frac{\ket{0}_c+\ket{1}_c}{\sqrt{2}}$
\State Define coin operator $\hat{C}_i = \hat{C}(T, \theta_i)$ for each site $i$ with $T=2$ (say)
\State Create topological phase boundary at site 0:

\textbf{       Ensure: } Rotation angle $\theta_0$ belongs to a topological phase e.g. see Fig.~2 in main.

\textbf{       Ensure: } Rotation angles $\theta_1=\theta_2=...=\theta_7$ ($\ne \theta_0$) belongs to a topological phase, e.g. see Fig.~2 in main.
\State Construct the global coin operator: $\hat{C} = \sum_{i=0}^{N-1} \ket{i}\bra{i} \otimes \hat{C}_i$
\State Construct shift operator $\hat{S}$ to implement cyclic movement on the $N$-cycle
\State Construct the evolution operator: $\hat{U} = \hat{S} \cdot \hat{C}$

\For{$t = 1$ to $J$}
    \State Update state: $\ket{\psi(t)} = \hat{U} \ket{\psi(t-1)}$
    \For{$x = 0$ to $N-1$}
        \State Calculate probability at site $x$: 
        $$P(x,t) = |\langle x,0|\psi(t)\rangle|^2 + |\langle x,1|\psi(t)\rangle|^2$$
        \State Store in the $(J+1)\times N$ array: $\texttt{pro}[t][x] \gets P(x,t)$
    \EndFor
\EndFor

\State \Return $\texttt{pro}$: probability distribution over time and space, i.e., probability $P(x)$ of finding the quantum walker at sites $x$, for different time $t$ as shown in Fig.~4 in main.
\end{algorithmic}

\subsubsection{Effects of dynamic disorder on generated edge states via step-dependent CQW on cyclic graphs}
Here, we provide our algorithm to implement the dynamic coin disorder in a CQW and its effect on the topological edge state in cyclic graphs (Sec.~D).
\begin{algorithmic}[1]
\Require \# of time steps $J$, \# of sites $N = 8$ (say), time-dependency parameter $T$
\Require  Coin parameters i.e., topological rotation angles $\theta_i$ which give non-zero winding numbers
\Require Number of disorder realizations $D = 500$ (say), disorder strength $\Delta_d=0.025$ (say)
\Ensure Probability distribution $\texttt{pro}[t][x]$ for $t=0$ to $J$, $x=0$ to $N-1$ for each disorder realization
\Ensure Averaged probability distribution $\texttt{avg\_pro}[t][x]$ for $t = 0$ to $J$, $x = 0$ to $N-1$ over $D$ disorder realizations

\State Initialize: $\texttt{avg\_pro} \gets$ array of zeros of shape $(J+1, N)$

\State Initialize the walker's state $\ket{\psi(0)}=\ket{0}\otimes\frac{\ket{0}_c+\ket{1}_c}{\sqrt{2}}$

\State Define coin operator $\hat{C}_i = \hat{C}(T, \theta_i)$ for each site $i$ with $T=2$ (say)
\State Construct shift operator $\hat{S}$ to implement cyclic movement on the $N$-cycle
\State Create topological phase boundary at site 0:

\textbf{       Ensure: } Rotation angle $\theta_0$ belongs to a topological phase e.g. see Fig.~2 in main.

\textbf{       Ensure: } Rotation angles $\theta_1=\theta_2=...=\theta_7$ ($\ne \theta_0$) belongs to a topological phase, e.g. see Fig.~2 in main.

\For{$s = 1$ to $D$} \Comment{Run over disorder realizations}

 \State \textbf{Generate} time-dependent random numbers $\delta \theta(t)$ of size $J$ via uniform distribution with disorder strength $\Delta_d$
\State \textbf{Ensure: } For each site $i \in [0,N-1]$, sample $\theta_i \rightarrow \theta_i +\Delta_d \delta \theta(t)$ for a particular time step $t$
    \State Construct the global coin operator: $\hat{C} = \sum_{i=0}^{N-1} \ket{i}\bra{i} \otimes \hat{C}_i$

\State Construct the evolution operator: $\hat{U} = \hat{S} \cdot \hat{C}$ 
    \For{$t = 1$ to $J$}
     \State $\theta_i \rightarrow \theta_i +\Delta_d \delta \theta(t)$ for each site $i$
        \State Evolve: $\ket{\psi(t)} = \hat{U} \ket{\psi(t-1)}$
        \For{$x = 0$ to $N-1$}
            \State Compute probability: $P(x,t) = |\langle x,0|\psi(t)\rangle|^2 + |\langle x,1|\psi(t)\rangle|^2$
            \State Update average: $\texttt{avg\_pro}[t][x] \gets \texttt{avg\_pro}[t][x] + P(x,t)$
        \EndFor
    \EndFor
\EndFor

\State Normalize: $\texttt{avg\_pro} \gets \texttt{avg\_pro} / D$
\State \Return $\texttt{avg\_pro}:$ probability distribution over time and space, i.e., probability $P(x)$ of finding the quantum walker at sites $x$, for different time $t$ considering the dynamic disorder effects, as shown in Fig.~\ref{edge-8-dyn} above.
\end{algorithmic}

Similarly, algorithms that implement static disorder and phase-preserving perturbations are also designed, following Sec.~D above. These algorithms are implementable in Python.
\subsection{Summary}
%In the main text, we propose CQW dynamics on finite cyclic graphs (lattices) with discrete Fourier transform and effective Hamiltonian, to simulate exotic topological effects. We show both step-dependent and step-independent CQWs offer flexible and resource-saving simulation platforms to generate topological phases (nonzero winding numbers), Dirac cones, topological gapped flat bands and edge states with excellent controllability via step-dependency, site number, periodic evolution and coin-rotation angles. Some 

Analytical derivations and numerical results supplementing the results and statements of the main text are provided in this supplementary material (SM). 

In SM Sec.~A, we detail the process of diagonalizing the translation and coin operators in momentum basis, and we then evaluated the energy dispersion, group velocity and effective mass of the quantum particle (quantum walker) evolving via CQW dynamics with finite cyclic graphs (lattices) taking recourse to discrete Fourier transform and unitary evolution. Then we derive the topological invariant: winding number for the cyclic graphs and explain how small cyclic graphs offer a resource-saving and flexible platform in contrast to infinite and multidimensional graphs, to simulate topological phenomena, including topological phases, flat bands, Dirac cones, and edge states. These small cyclic graphs offer excellent controllability over these topological effects via CQW parameters: step-dependency, site number, periodic evolution and coin-rotation angles. Numerical results for energy dispersion and topological phases characterized by winding numbers, in 3,4,7,8-cycles for both step-dependent and step-independent CQWs. 

We observe that odd and even cyclic graphs show distinct features in energy dispersion, and we prove that rotational flat bands are solely seen in even $4n$-cycles ($n\in\mathbf{N}$) in SM Sec.~B. Moreover, we prove that energy gap closing in rotation space implies energy gap-closing (Dirac cones) in momentum space. We further show the generation of topological gapped flat bands, and we derive the condition to obtain these flat bands, which are verified with zero group velocity and undefined effective mass. In SM Sec.~C, we establish how to generate topological edge states at the interface between two distinct topological phases with both odd and even cycle graphs of finite size. Obviating the need for resource-consuming models like split-step or split-coin quantum walks (QWs), in order to generate edge states in real physical systems (e.g., photonic or electronic) and from the use of infinite or multi-dimensional lattices, are advantages of our CQW setup with small cyclic graphs. This facilitates the most resource-saving and straightforward practical implementations of our scheme in physical platforms such as photonic or electronic systems, to design and control topological phenomena.

In Sec.~D, we numerically demonstrate that the generated topological edge states via our CQW dynamics on small cyclic graphs are robust against static and dynamic disorder (introduced through gate/coin operations), as well as robust against phase-preserving perturbations. 
This makes the topological phases and their protected edge states generated via our CQW scheme potentially useful for noise-resilient quantum information processing and fault-tolerant quantum computing. In Sec.~E, we demonstrate that our step-dependent CQW (SD-CQW) scheme for generating edge states is state-independent, i.e., it is completely independent of the nature (superposed, out-of-phase superposed, uneven superposed and unsuperposed) of the initial coin state or the initial state of the quantum walker. We compute the resource overhead (operator count plus detector count) of our SD-CQW scheme in Sec.~F.  Our protocol requires $G_{\rm SD} = 2\tau$ operators, two distinct coins (for phase boundary), and a constant number of detectors $N$ (e.g., $N=7,8$), whereas existing split-step and split-coin schemes require $G_{\rm SS/SC} = 4\tau$ operators, three distinct coins (for phase boundary), and $(2\tau+1)$ detectors, which scale linearly with the number of time steps $\tau$. Thus, our scheme greatly enhances experimental feasibility and efficiency.
Finally, in Sec.~G, we provide algorithms and Python code in GitHub~\cite{git} to generate edge states and simulate disorder effects and effects of different initial states on the edge states within our SD-CQW (with cyclic graphs) framework.

\twocolumngrid

\end{document}